\documentclass{aa}
\usepackage{graphicx}
\usepackage{txfonts}

\newcommand{\kms}{\,km\,s$^{-1}$}
\newcommand{\ang}{\,\AA}

\newcommand{\Mg}{Mg{\sc\,ii}}
\newcommand{\Cb}{C{\sc\,ii}}
\newcommand{\Si}{Si{\sc\,iv}}
\newcommand{\Cl}{Cl\,{\sc i}}

\usepackage{makeidx}
\makeindex

\begin{document} 

\authorrunning{C. E. Alissandrakis et al.}
\title{Evidence for two-loop interaction from IRIS and SDO observations of penumbral brightenings
}
\author{C. E. Alissandrakis, A. Koukras, S. Patsourakos \and A. Nindos
}

\institute{Department of Physics, University of Ioannina, GR-45110 Ioannina, 
Greece\\
\email{calissan@cc.uoi.gr}
}

\date{Received ...; accepted ...}

  \abstract
 {}
{We investigate small scale energy release events which can provide clues on the heating mechanism of the solar corona.}
{We analyzed spectral and imaging data from the Interface Region Imaging Spectrograph (IRIS), images from the Atmospheric Imaging Assembly (AIA) aboard the Solar Dynamics Observatoty (SDO), and magnetograms from the Helioseismic and Magnetic Imager (HMI) aboard SDO.
} 
{We report observations of small flaring loops in the penumbra of a large sunspot on July 19, 2013. Our main event consisted of a loop spanning $\sim15$\arcsec, from the umbral-penumbral boundary to an opposite polarity region outside the penumbra. It lasted approximately 10 minutes with a two minute impulsive peak and was observed in all AIA/SDO channels, while the IRIS slit was located near its penumbral footpoint. Mass motions with an apparent velocity of $\sim100$\kms\ were detected beyond the brightening, starting in the rise phase of the impulsive peak; these were apparently associated with a higher-lying loop.  We interpret these motions in terms of two-loop interaction.  IRIS spectra in both the {C\,\sc ii} and {Si \sc iv} lines showed very extended wings, up to about 400\,km/s, first in the blue (upflows) and subsequently in the red wing. In addition to the strong lines, emission was detected in the weak lines of {Cl\,\sc i}, {O\,\sc i} and {C\,\sc i}, as well as in the {Mg\,\sc ii} triplet lines. Absorption features in the profiles of the {C\,\sc ii} doublet, the {Si\,\sc iv} doublet and the \Mg\ h and k lines indicate the existence of material with a lower source function between the brightening and the observer. We attribute this absorption to the higher loop and this adds further credibility to the two-loop interaction hypothesis. Tilts were detected in the absorption spectra, as well as in the spectra of {Cl\,\sc i}, {O\,\sc i,} and {C\,\sc i} lines, possibly indicating rotational motions from the untwisting of magnetic flux tubes.}
{We conclude that the absorption features in the \Cb, \Si\ and \Mg\ profiles originate in a higher-lying, descending loop; as this approached the already activated lower-lying loop, their interaction gave rise to the impulsive peak, the very broad line profiles and the mass motions.}


   \maketitle

\section{Introduction}
Energy release on the Sun occurs on a large range of spatial and temporal scales, from huge flares and Coronal Mass Ejections (CMEs) to tiny events barely resolved by our most advanced instruments. Large scale events, important as they might be due to their effects in the interplanetary space and on Earth, are usually too complex to analyze in detail and reveal the elementary physical processes involved. Moreover, these events can be considered as the non-linear interaction of small-scale energy release incidents \citep[see, e.g.,][]{1994SSRv...68...39V, 2012SSRv..173..223C}, an approach which has led to the concept of fragmentation of energy release.

The interaction of two adjacent  current-carrying magnetic loops of equal length with magnetic fields in the opposite direction was one of the first models proposed to explain solar flares \citep{1960MNRAS.120...89G}. Several extensions and variants of this essentially quadrupolar two-loop interaction model have emerged \citep[e.g.,][]{1977ApJ...216..123H, 1995ApJ...450..422K, 1997ApJ...489..976N, 1999PASJ...51..483H, 1999PASJ...51..553U, 2014ApJ...786L..21A}. The two loops could come close enough to interact by photospheric shearing motions applied to their footpoints or by the emergence  of one of them in the vicinity of the other. Depending on various parameters such as the relative angle between the two interacting flux tubes and their  magnetic helicity sign, different evolutionary paths, including flux-tube bouncing, merging, slingshot and tunneling could arise \citep{2001ApJ...553..905L}. For example, extreme ultraviolet (EUV) spectral and  imaging observations of two interacting cool loops reported by Li et al. (2014) revealed the transformation of the mutual helicity of the system into self-helicity. Observations, mainly in the X-rays but sometimes also in EUV and radio, showed that two-loop interactions could be relevant to a fraction of observed solar flares \citep[e.g.,][]{1988ApJ...326..425M, 1993A&A...271..292D, 1999ApJ...526.1026A, 2004A&A...420..351K, 2010ApJ...723.1651K,2013ApJ...764L...3C}. Two-loop interactions could be also found at smaller spatial scales than those of flares, that is, in small-scale active region brightenings \citep[e.g.,][]{1994ApJ...422..906S}.

Small scale events occur all the time and everywhere, in active regions as well as in the so-called quiet sun. Every time a new instrument with increased spatial and temporal resolution appears, more varieties of such events are added to a long list and recent instruments are no exception, having provided material for a large number of publications \citep[e.g.,][among others]{2014Sci...346C.315P, 2014ApJ...790L..29T,  2014ApJ...797...88H, 2016ApJ...816...92T, 2016ApJ...823...60B, 2016ApJ...829...35W, 2016ApJ...829..103D, 2016ApJ...833...22C, 2016A&A...592A.100R}. However, we are probably still quite far from the nanoflare limit considered by many to be the origin of the coronal heating \citep{1983ApJ...264..642P, 1988ApJ...330..474P}. In any case, small scale events, being relatively simpler in their spatial structure and time evolution than their large-scale counterparts, provide a good opportunity to check existing models of energy release and atmospheric structure and to develop new models \citep[e.g.][among others]{2006SoPh..234...41K, 2008ApJ...687.1363B, 2010ApJ...718..981R,2014Sci...346B.315T, 2016SSRv..201....1R}.

In a previous publication \citep{2013A&A...556A..79A} we reported on the spatial and temporal characteristics of small brightenings observed with the Atmospheric Imaging Assembly (AIA) in the umbrae of large spots; these turned out to be footpoints of hot flaring loops which had their other end in opposite magnetic polarity regions  44 to 53\arcsec\ away. Subsequently, we conducted a systematic search for similar events, by inspecting AIA images in the 1600\,\AA\ band from the {\it helioviewer} site.\footnote{https://helioviewer.org/.} This revealed a large number of brightenings in sunspot umbrae and penumbrae. 

Here we report on a particular penumbral brightening that occurred on July 19, 2013, which attracted our attention both because of its morphology and because of the existence of simultaneous observations with the Interface Region Imaging Spectrograph (IRIS). While processing  the IRIS data, we detected more such events that were too faint to notice in the Solar Dynamics Observatory (SDO) data and we found evidence of interaction between a low-lying loop and a higher one. In Section 2 we describe the observations and their analysis, in Sections 3 and 4 we present the results and in Section 5 the conclusions.

\section{Observations and analysis}
\begin{figure}[h!]
\centering
\includegraphics[width=\hsize]{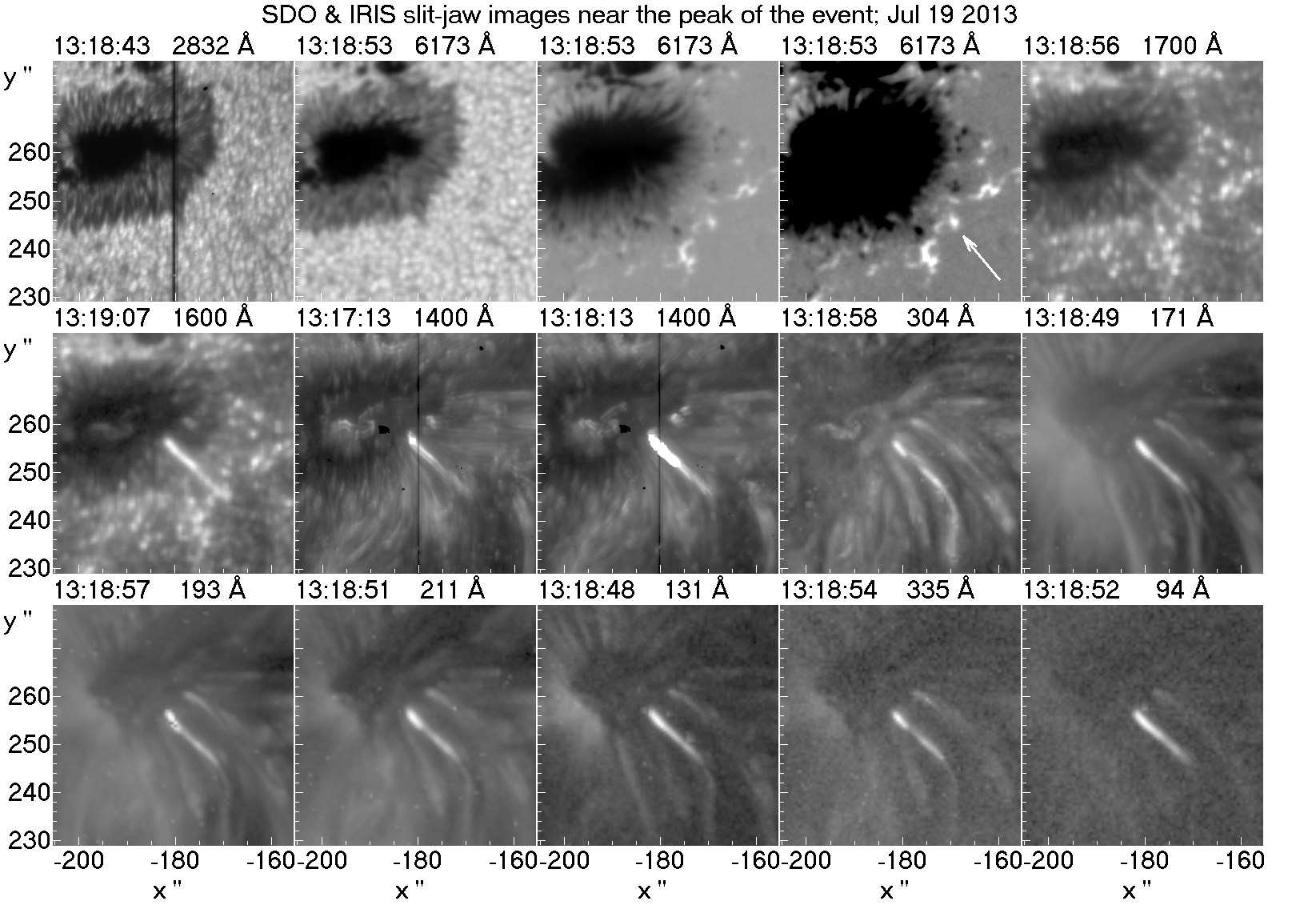}
\caption{Images near the maximum of the main event: Continuum,  longitudinal magnetic field (full range and saturated at $\pm300$\,G) from HMI, IRIS slit-jaw (SJ) images in the 2832 and 1400\,\AA\ bands and all AIA channels. The AIA 171 and 193 as well as the 1400 IRIS SJ images have saturated pixels near the peak; for this reason we give two 1400 images. The arrow points to a magnetic feature discussed in the text.
}
\label{overview}
\end{figure}

The main event occurred around 13:18 UT of July 19, 2013, at E11.8 S20.4, in National Oceanic and Atmospheric Administration (NOAA) Active region 11793. Images near the peak of the event are presented in Fig. \ref{overview}. The IRIS observing sequence started at 12:01 UT and ended at 13:41, thus we used AIA and Helioseismic and Magnetic Imager (HMI) data for the same time interval; we retrieved IRIS level 2 data from the IRIS site and AIA level 1.0 data from the Joint Science Operations Center (JSOC) site.

\subsection{Corrections to SDO observations}
Although the pointing in individual AIA wavelength bands is very stable, we noticed slight pointing differences among bands that needed correction. We first determined the correction for the AIA 1600 and 1700\,\AA\ bands by comparing them with the absolute value of the HMI longitudinal magnetic field (outside spots), which proved better in this respect than the HMI continuum; subsequently we computed the correction for the 304\,\AA\ band with respect to the 1700\,\AA\ images and then for the other AIA bands with respect to 304. The values of the pointing offset ranged from 0.7 to 1.6 pixels in the EW direction and from  $-1$ to 1 pixels in the NS direction.

\begin{figure}[h!]
\centering
\includegraphics[width=\hsize]{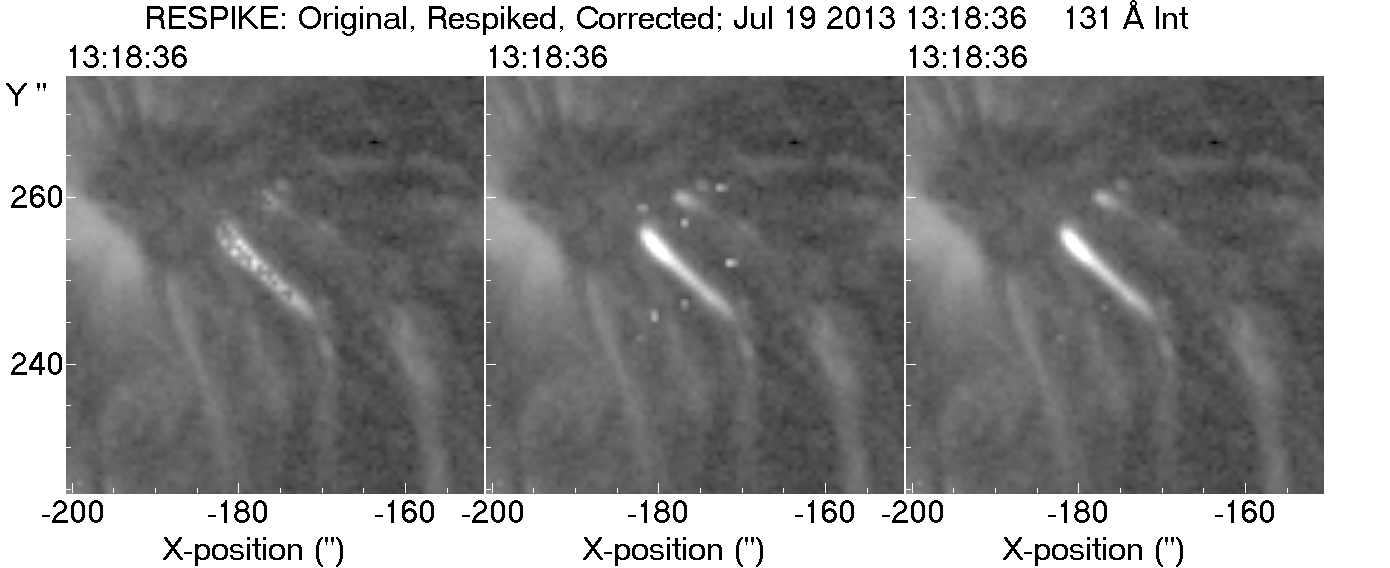}
\caption{Respike correction. Left: Image in the 131\,\AA\ band, corrected for hot pixels by the AIA algorithm. Center: Reconstituted original image. Right: Final image after deletion of the real hot pixels.
}
\label{respike}
\end{figure}

We found that the AIA algorithm that removed hot pixels was not working properly and, as a result, some images showed dark pixels where there should not be any (see left panel in Fig. \ref{respike}); this occurred in high intensity compact structures near the maximum of the event. In order to correct for this effect, we retrieved the original values of the corrected pixels from the AIA database and reconstituted the original image (central panel in Fig. \ref{respike}), from which we removed the real hot pixels manually. We did this for a total of 307 images (out of 3500) in the AIA EUV channels; no corrections were necessary for the AIA UV channels.

\begin{figure}[h!]
\centering
\includegraphics[width=\hsize]{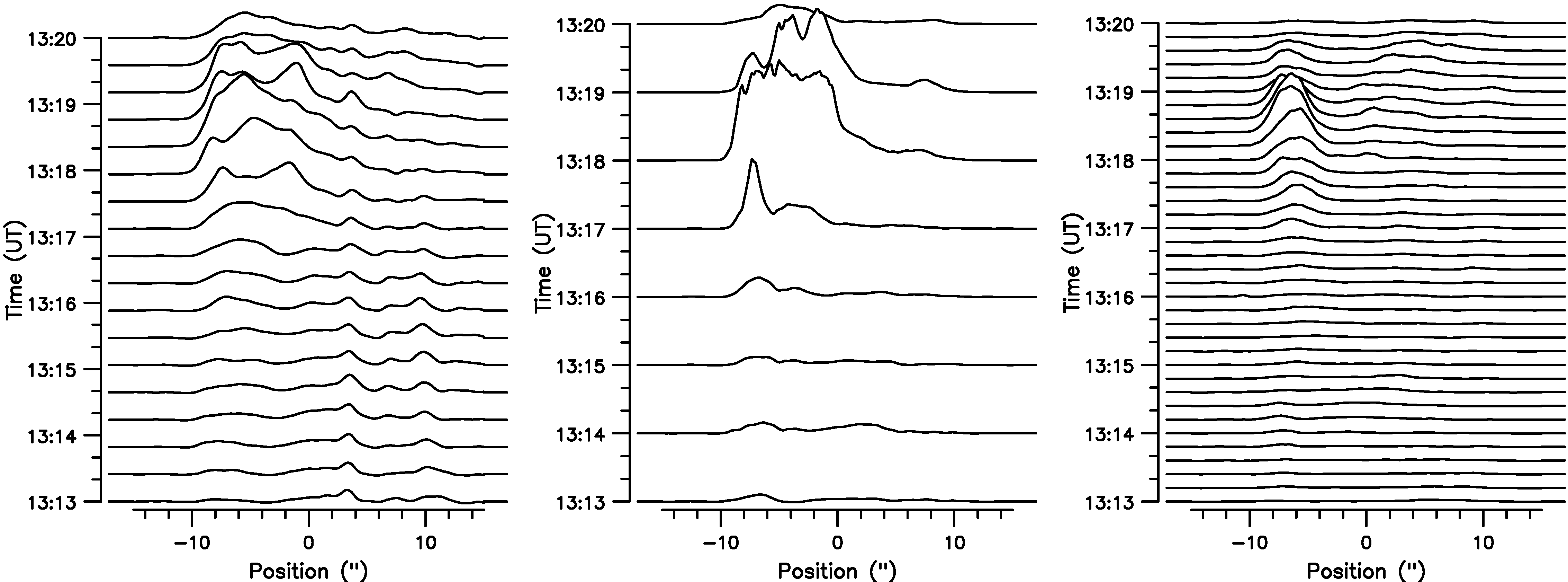}
\includegraphics[width=\hsize]{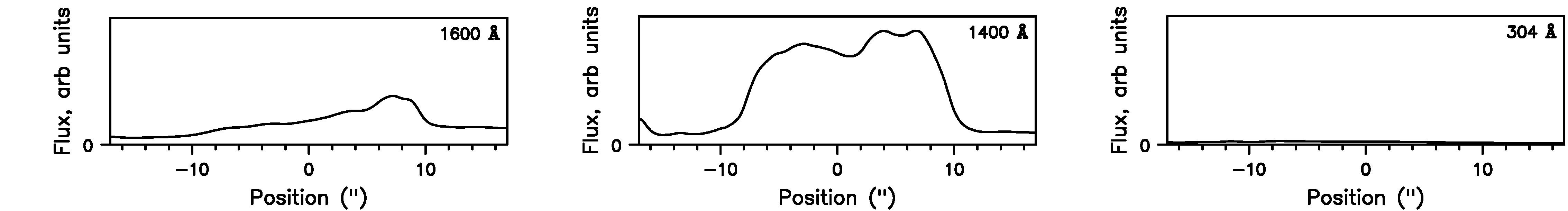}
\caption{Intensity along the loop integrated in the perpendicular direction as a function of time for the main event at 1600 (left), 1400 (center) and 304\ang\ (right). The time average, shown in the bottom frames, has been subtracted. At 1400\,\AA, many points in the 13:18 and some in the 13:17 and 13:19 UT tracings are saturated. Zero position is near the center of the loop.}
\label{cuts}
\end{figure}

\begin{figure*}[ht]
\centering
\includegraphics[width=\textwidth]{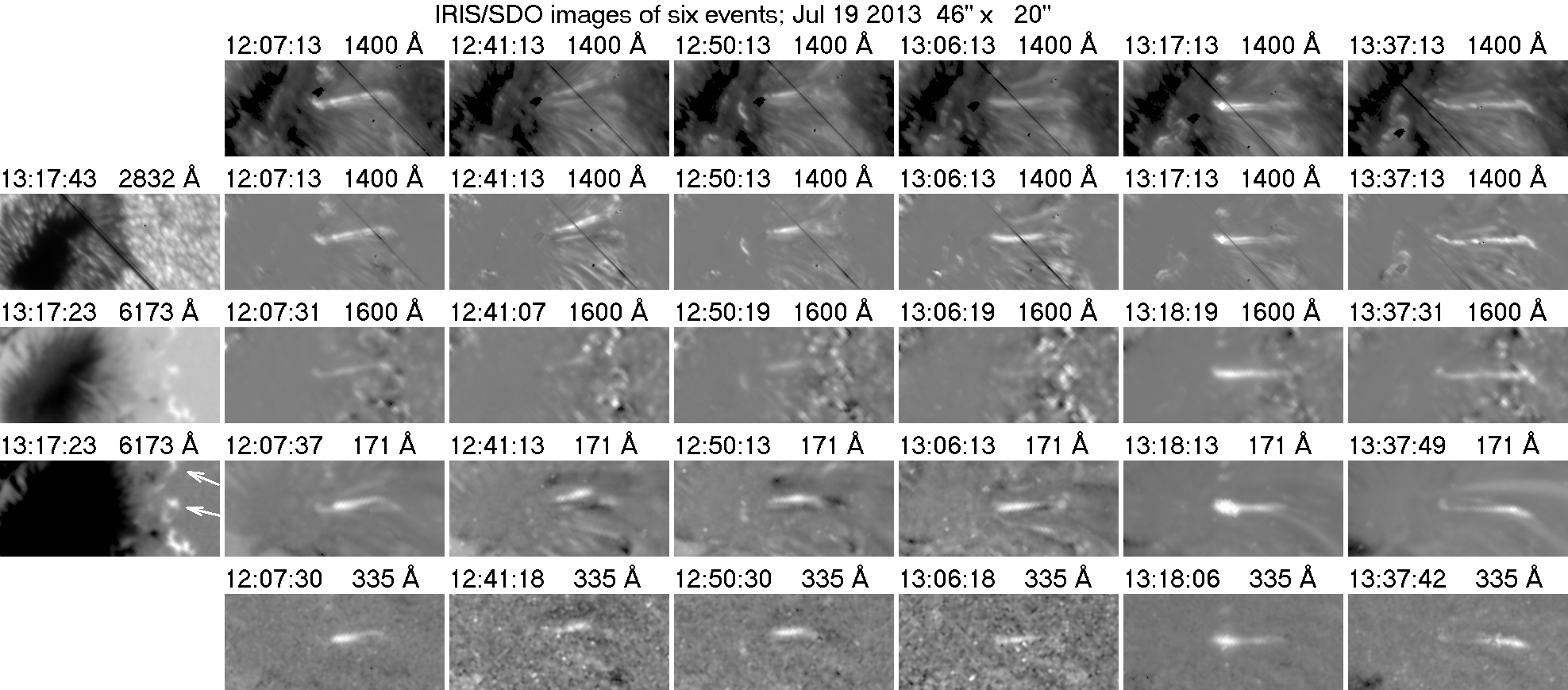}
\caption{Images of penumbral brightenings in the IRIS 1400\,\AA\ SJ band (top row). The other rows show difference images in the 1400, 1600, 171 and 335\,\AA\ bands, normalized  to the minimum and maximum intensity for clarity. In the left column we give a 2832\,\AA\ continuum image, as well as full-range and saturated magnetograms for reference. The images have been rotated by 44\degr\ counter-clockwise. 
}
\label{others}
\end{figure*}

\subsection{IRIS data and corrections}
The IRIS observing sequence contains 200 spectra and 200 slit-jaw (SJ) images. Spectra were taken every 30\,s and we have the full near UV (NUV) and far UV (FUV) bands, which include the \Mg\ h and k lines near 2800\,\AA, the \Cb\ doublet at 1335\,\AA\ and {Si \sc iv} doublet near 1400\,\AA. The sampling step in the direction of dispersion was 0.025\,\AA, which corresponds to 5.8\,km\,s$^{-1}$ in the FUV band and 2.7\,km\,s$^{-1}$ in the NUV band of IRIS; we note that the resolution of the spectrograph is 0.026\,\AA\ and  0.053\,\AA\ respectively \citep{2014SoPh..289.2733D}. Along the slit the sampling step was 0.17\arcsec, which is about two times smaller than the effective spatial resolution of IRIS ($\sim0.4$\arcsec). We thus have about two pixels per resolution element, both in the spatial and the spectral direction. The slit was oriented in the solar NS direction, its width was 0.33\arcsec\ and the exposure time was 10\,s.

Slit-jaw images were taken in the 1400\,\AA\ band (of 55\,\AA\ width, containing the {Si\,\sc iv} doublet) simultaneously with the odd-numbered spectra in the sequence  and in the 2832\,\AA\ band (of 4\,\AA\ width, in the quasi-continuum in the far wing of \Mg\ h line) simultaneously with the even-numbered spectra; we thus have 100 SJ images in each band, with a cadence of 1\,min. The exposure time of the SJ images was also 10\,s.

Our events occurred during the early days of IRIS operation and, as a consequence, the pointing was not stable; the overall drift was 25 pixels in the EW direction and nine pixels in the NS direction. We corrected the pointing of the SJ images by measuring the displacement of a 1400 image near the event with respect to the corresponding AIA 1600\,\AA\ image and then corrected all 1400 images with respect to that. We did the same with the 2832\,\AA\ SJs, this time using HMI continuum images as a reference. SJ movies created after the corrections show no appreciable jitter, thus we estimate the correction to be good within about 0.5 pixel. Another problem was image motion along the slit during the exposure (see, e.g. the 1400-band frames in Fig. \ref{overview}), making some features appear double; this happened only for the 1400 SJ and the spectra taken simultaneously.
In the movie attached to Fig. \ref{movie01} we present both sets of SJ images after jitter corrections; this gives an overview of the activity, the quality of the images and the efficiency of the corrections
 
Pointing corrections for spectra were more complicated, due to the fact that for each spectrum we have an SJ image in one band only, 1400 or 2832. We corrected the NS pointing using the fiducial marks, by comparing the odd-numbered spectra with the corresponding 1400-band SJs and even-numbered spectra with the 2832-band SJs. We found a displacement among even and odd spectra, which required an additional pointing correction in the NS direction to the 1400 SJ images and spectra.

\section{Results from imaging data}

\subsection{Morphology}
The main event at 13:18 UT (Fig. \ref{overview}) appears as a loop, with its NE footpoint  very close to the umbra-penumbra boundary and its SW footpoint near a magnetic patch of opposite (positive) polarity (marked by an arrow in the figure), 8\arcsec\ from the penumbra-photosphere border. It was fairly bright, which caused several pixels in the IRIS 1400\,\AA\ and the AIA 171 and 193\,\AA\ AIA images to saturate. 

The appearance of the event was very similar in all wavelength bands, even in the 1700\,\AA\ channel; the AIA image that resembles most the IRIS 1400 image is the one at 1600\,\AA, indicating that these bands sample similar physical conditions, at least in this case. The structure had a length of 17\arcsec\ with a very bright 9\arcsec\ long section extending from the inner (NE) footpoint to about the penumbra-photosphere boundary. This bright section was more compact in the 304\,\AA\ and higher temperature bands, where its length was about 5\arcsec. The width of the loop in the IRIS 1400 SJs was of the order of 0.5\arcsec, much too small for AIA to resolve.

In addition to this, other inhomogeneities appeared along the loop, best seen in the 1400\,\AA\ IRIS SJs due to their high resolution, with some of them marginally detectable in the AIA images. This is illustrated in Fig. \ref{cuts}, which shows the intensity variation along the loop, after subtraction of the average, at several instances during the event in the 1600, 1400 and 304\,\AA\ bands. In order to avoid transient brightenings, the average intensity was calculated by computing the histogram of values at each pixel and rejecting the top 10\%. Several bright points of 1-2\arcsec\ size are visible along the loop at 1600 and 1400\,\AA; we note that the peaks at 3.7 and 10\arcsec\ in the 1600 plots are not related to the event.

The situation is similar with the other brightenings detected during the IRIS observing sequence. The strongest of them are displayed in Fig. \ref{others}, together with the main event. The figure shows images in the 1400, 1600, 171 and 94\,\AA\ bands, which cover a temperature range from the chromosphere to the hot corona. Images at 1600, 171 and 335\,\AA\ are shown after subtraction of the time averages, 1400\,\AA\ images are given before and after subtraction, while a 2832\,\AA\ band image and a magnetogram have been added for reference. We will number these events from 1 to 6, in order of their appearance, the main event being number 5. In addition to the main event, other events such as those at 12:07, 12:41 and 13:37 UT show prominent small-scale structures along their length in the 1400\,\AA\ band. These are also detectable in the 1600\ang\ band, despite the lower resolution, while in the hotter AIA bands the intensity along the brightenings varies more smoothly.

\begin{figure}[ht]
\centering
\includegraphics[width=.9\hsize]{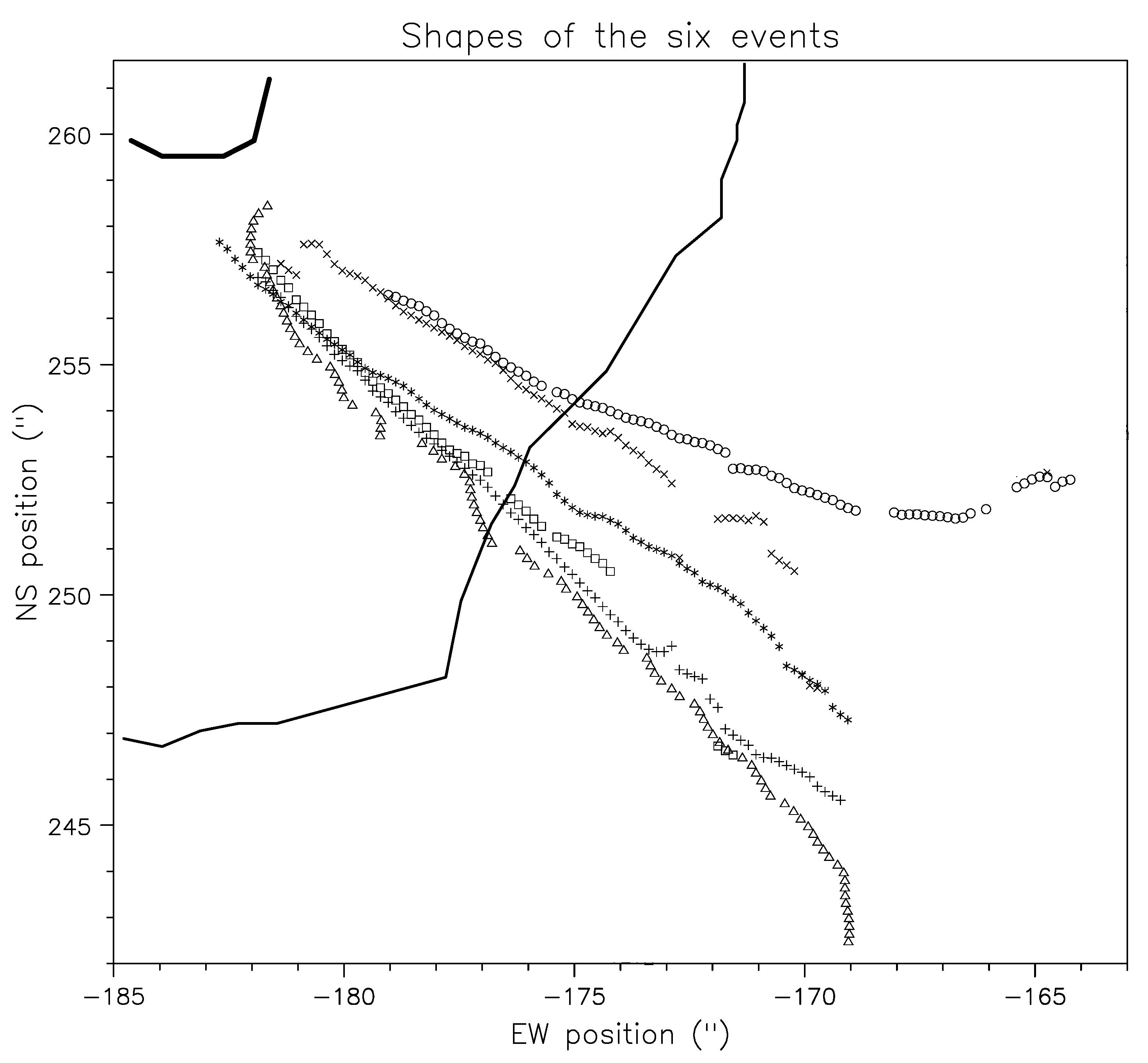}
\caption{Shapes of the six events, determined from the position of the maximum intensity of their average SJ images in the 1400\,\AA\ band and plotted in the EW-NS coordinate system. Asterisks refer to event 1, circles to event 2, x's to event 3, squares to event 4, pluses to event 5 (the main event) and triangles to event 6. Full lines show the penumbra-chromosphere (thin) and the umbra-penumbra (thick) boundaries, measured from the gradient of the continuum images.}
\label{shapes}
\end{figure}

It is interesting to compare the shapes and positions of the penumbral brightenings. This can be done better with the help of Fig. \ref{shapes}, where we plotted the position of maximum intensity along the brightenings computed from the time-average image of each event in the 1400\,\AA\ SJ images; the positions are corrected for solar rotation and refer to 13:18 UT. We note that events 4, 5 and 6 are practically cospatial, with event 1 forming a small angle with respect to them. Events 2 and 3 are at a distance from events 4-6 and their direction forms an angle greater than that of event 1. Their outer footpoints appear to be associated with a different positive polarity patch than the other events; both patches are marked with arrows in Fig. \ref{others}. The inner footpoints of all events are very close to one another, near the umbra-penumbra boundary as already mentioned for the main event. Their projections on the plane of the sky are close to straight lines, with events 2 and 3 showing some curvature and the others bending to an ``S'' shape near their edges. All together, they give the impression of magnetic lines of force fanning out from the sunspot to the surrounding area.

An important question is whether there is anything peculiar with the magnetic field in the regions near the events. Inspection of the HMI magnetograms did not show anything noticeable, apart from some moving magnetic features all around the spot; there is, however, a difference between the orientation of the brightenings and that of penumbral filaments (Fig.  \ref{others}), as well as with the orientation of a low intensity large-scale loop visible in the difference images of events 5 and 6 at 171\,\AA. This is probably an indication of shear in the vertical direction.

\begin{figure}[h]
\centering
\includegraphics[width=0.9\hsize]{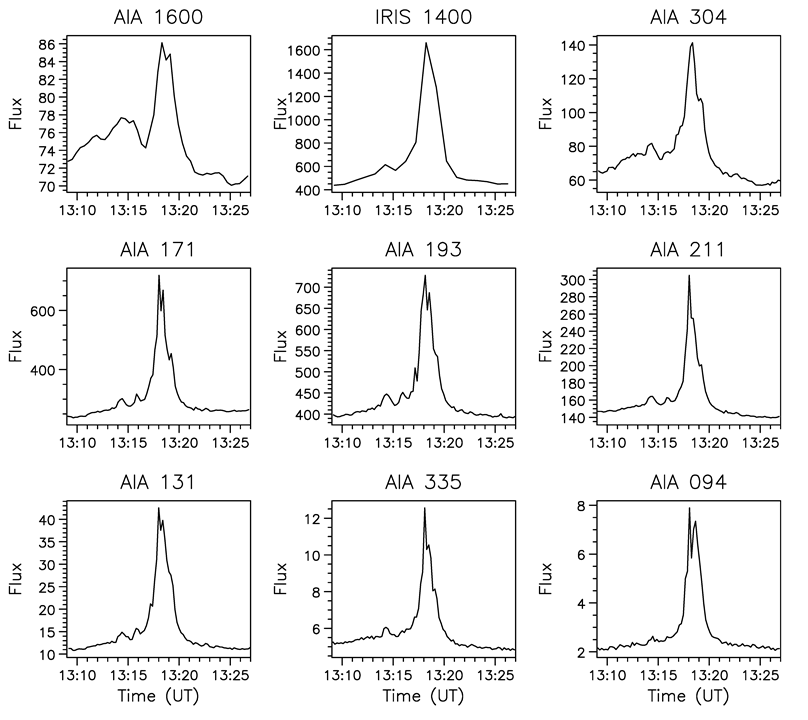}
\includegraphics[width=0.84\hsize]{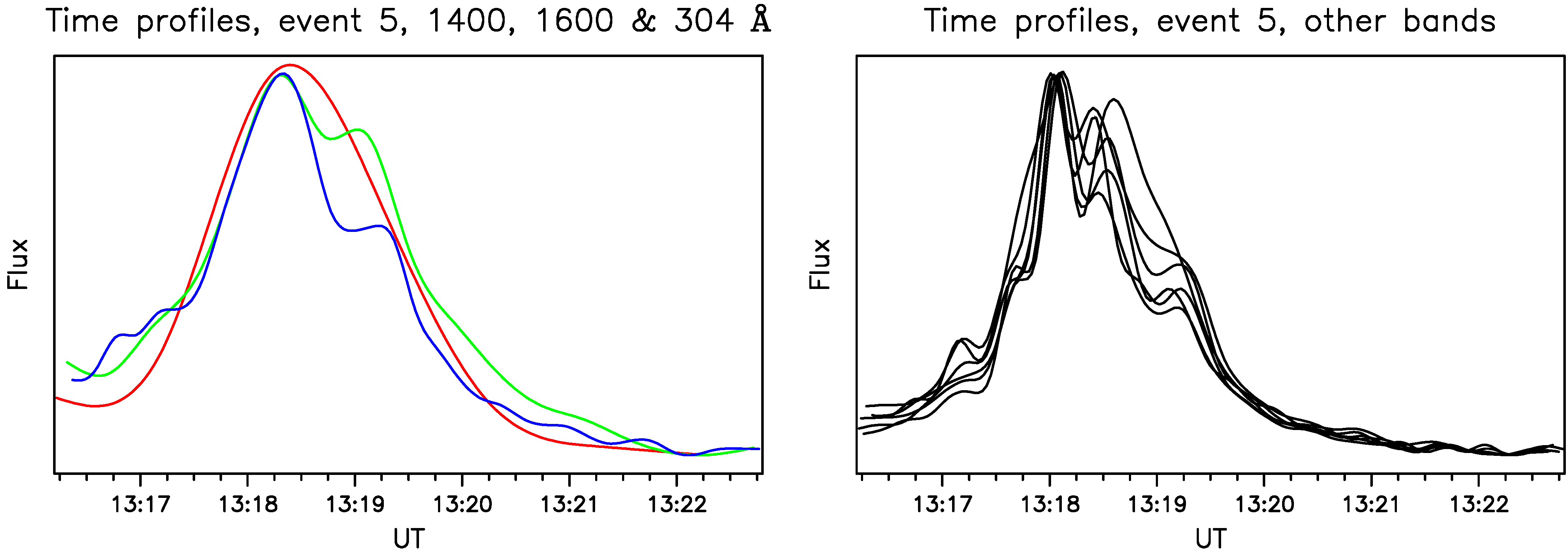}
\caption{Time profiles of the main event in all wavelength bands (top three rows). The bottom row shows time profiles near the spike in the 1600, 1400 and 304\,\AA\ bands (left) and all other bands (right), plotted together and normalized to their respective minima and maxima. Red is for 1400, green for 1600 and blue for 304\,\AA.
}
\label{timprof}
\end{figure}

\subsection{Time evolution}
All events were very short, as can be seen in Fig. \ref{timprof} which shows the time profile of the flux of the main event in all spectral bands. Contrary to other small events that we observed in the past \citep{2013A&A...556A..79A,2015A&A...582A..52A}, where the impulsive peak was limited near the loop footpoints and was followed by a ``post-burst'' phase associated with the body of the loop \citep[see Fig. 4 of][]{2015A&A...582A..52A}, this one showed just a spike-like peak. The duration of the spike was about 2 min in full width at half maximum (FWHM) and it was superposed on a slowly varying background that lasted about 10 min. As a matter of fact, the spike consisted of 2-3 components, as evidenced from the AIA images that have a better cadence (12\,s) than the IRIS SJ images (60\,s); it even appears that the AIA cadence is not sufficiently short to reveal the full temporal structure.

\begin{figure}
\centering
\includegraphics[width=.9\hsize]{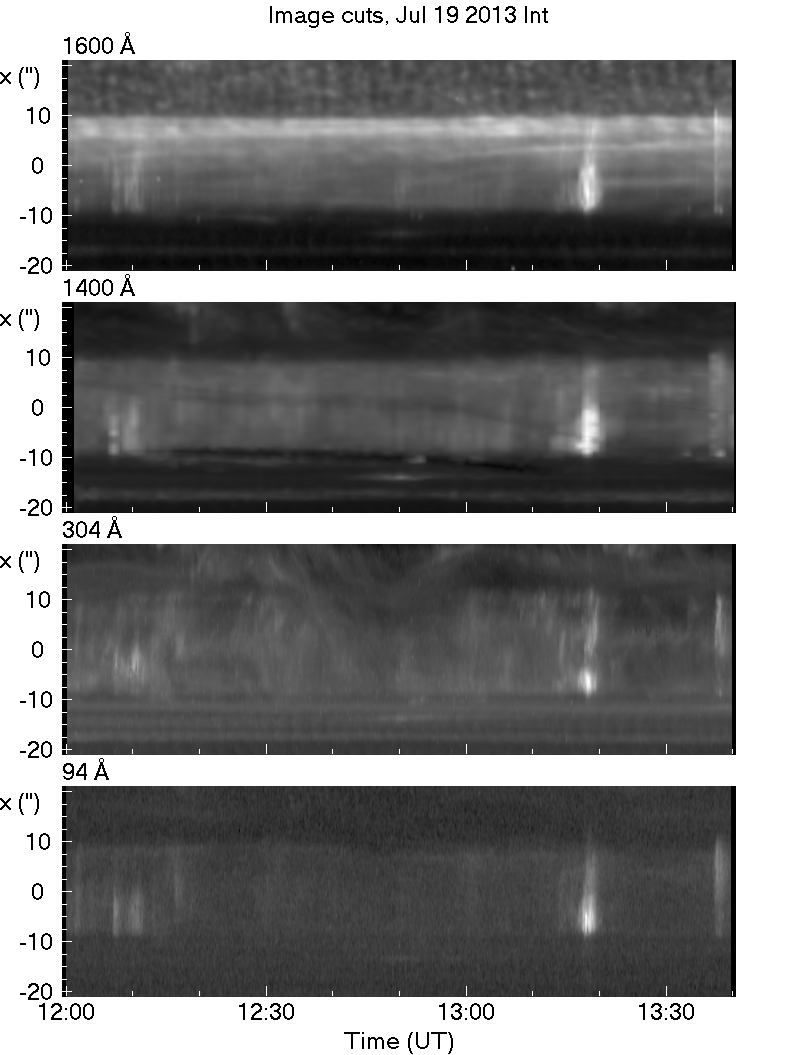}
\caption{Image of intensity as a function of time and position along the loop for the entire observing sequence in four wavelength bands. Zero position is near the middle of the brightenings, x increases away from the spot. The spectrograph slit appears as a dark band in the 1400\ang\ image, shifting from $x=6$ to $x=-10$ during the sequence.}
\label{cutsA}
\end{figure}

Although the time profiles were very similar in all bands, those in low temperature bands (1600, 1400 and 304\,\AA) had some differences from the high temperature band profiles (bottom plots of Fig. \ref{timprof}). The peaks of the latter precede the former by $\sim20$\,s but, for the bulk of the peak emission, cross-correlation gives a shift of $\sim7$\,s. We note that these values are smaller than the cadence of the 1600\ang\ and 1400\ang\ images, hence they should be considered as indicative only. Similar results were obtained for events 1 and 6, for which reliable time profiles could be computed.

A more complete view of the time evolution is given in Fig. \ref{cutsA}, where we present, in image form, the intensity integrated perpendicularly to the structures over 3\arcsec, as a function of position and time for the 1600, 1400, 304 and 94\,\AA\ bands. The integration was performed around the location of event 5, thus some of the other events are not clearly visible. An interesting feature is that the brightenings occurred in an area which was brighter than average throughout the observing session ({\it cf.} bottom plots of Fig. \ref{cuts}); as a matter of fact, the loop associated with event 5 is clearly visible in the minimum envelope of all images in the time sequence, in all wavelength bands. This shows that there was continuous low-level activity there.

\begin{figure}
\centering
\includegraphics[width=.95\hsize]{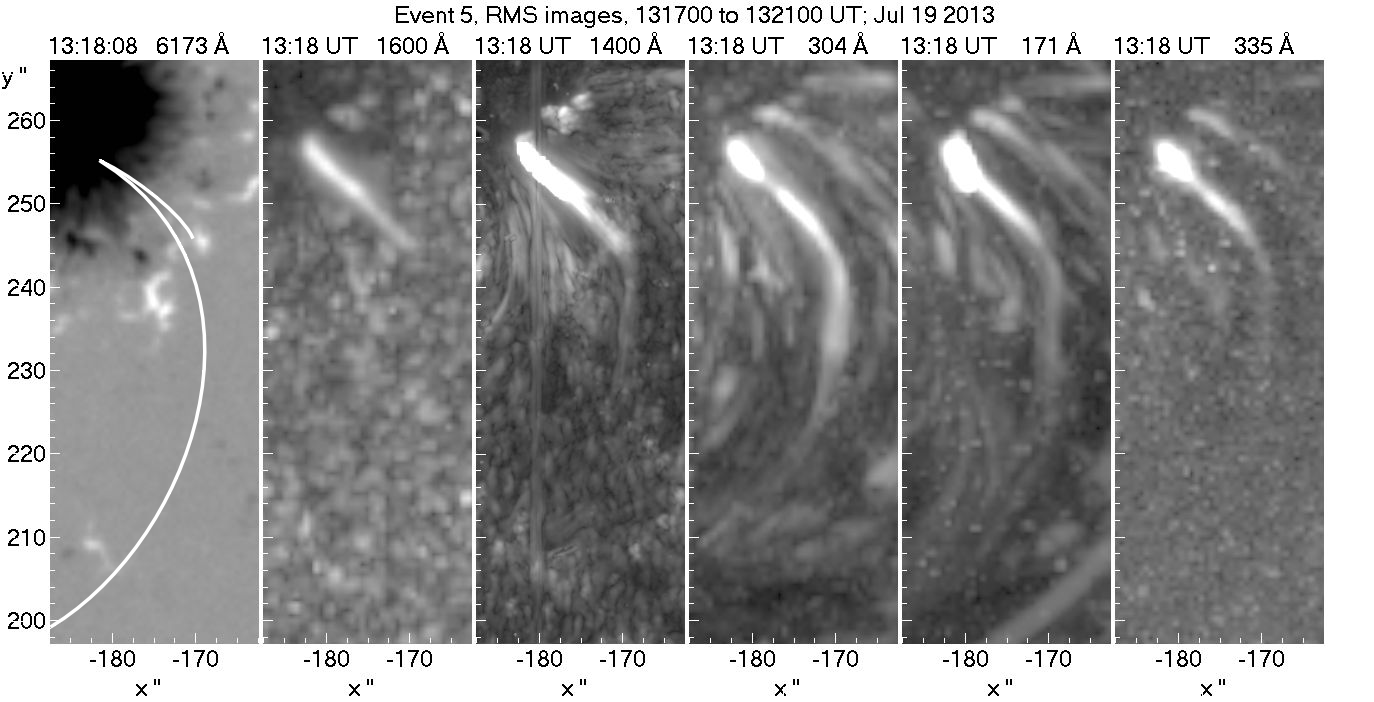}
\caption{HMI magnetogram and images of the root mean square ( RMS) of the intensity in five wavelength bands. High intensity regions in the 1400, 304, 171 and 335\ang\ images have been saturated to show better the low intensity regions associated with the mass flows. White lines are magnetic field lines of force (see text).
}
\label{flows}
\end{figure}

\begin{figure}[h]
\centering
\includegraphics[width=.9\hsize]{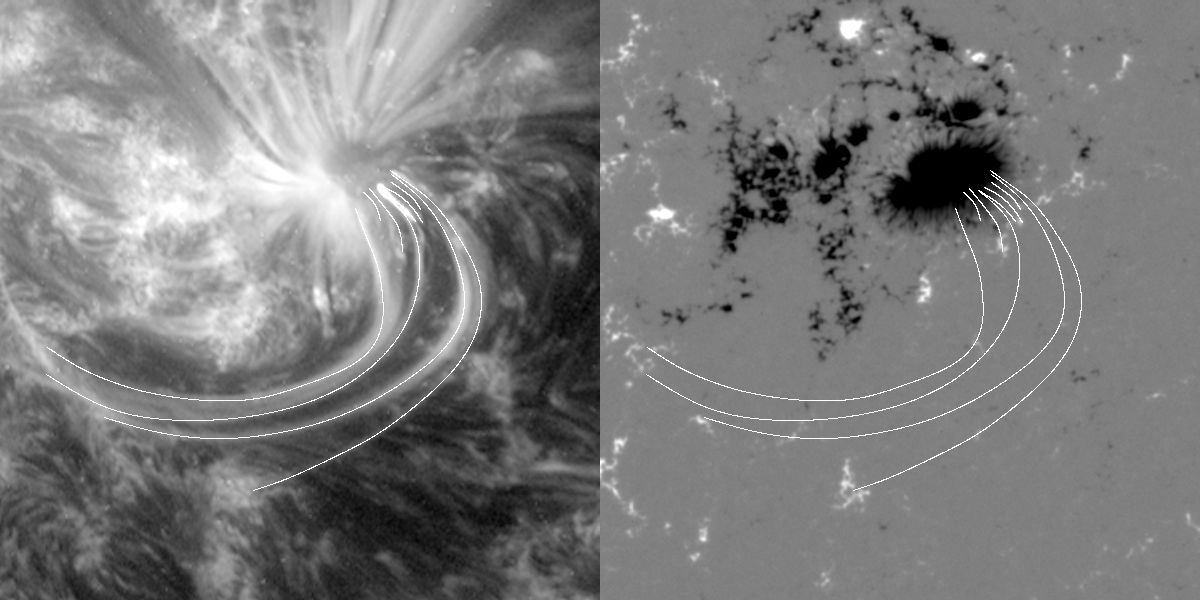}
\caption{Lines of force of the magnetic field on top of a 171\ang\ AIA image and a HMI magnetogram. The size of the region is 180 by 180\arcsec.
}
\label{FieldLines}
\end{figure}

\begin{figure}
\centering
\includegraphics[width=.9\hsize]{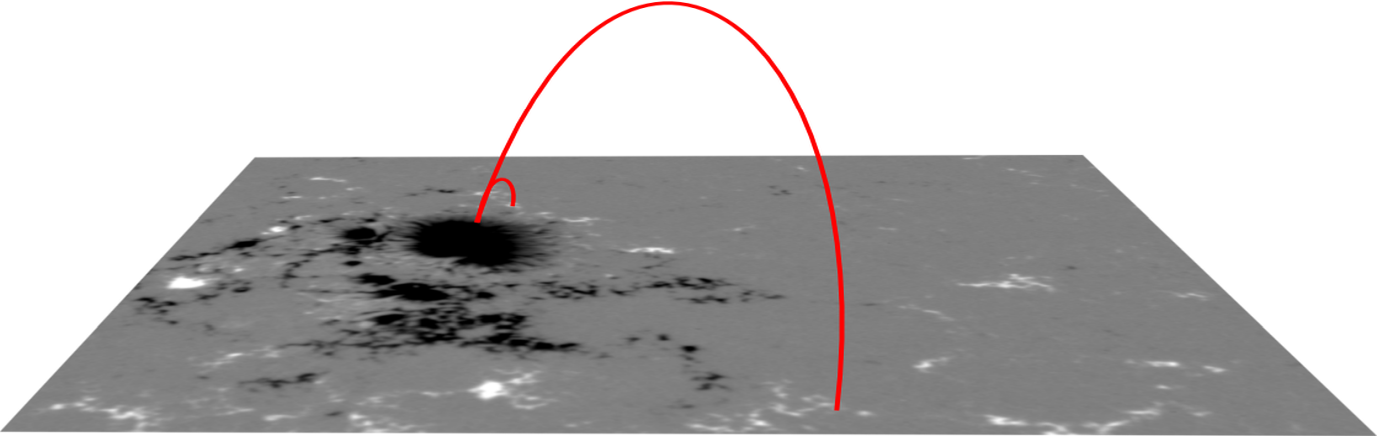}
\includegraphics[width=.9\hsize]{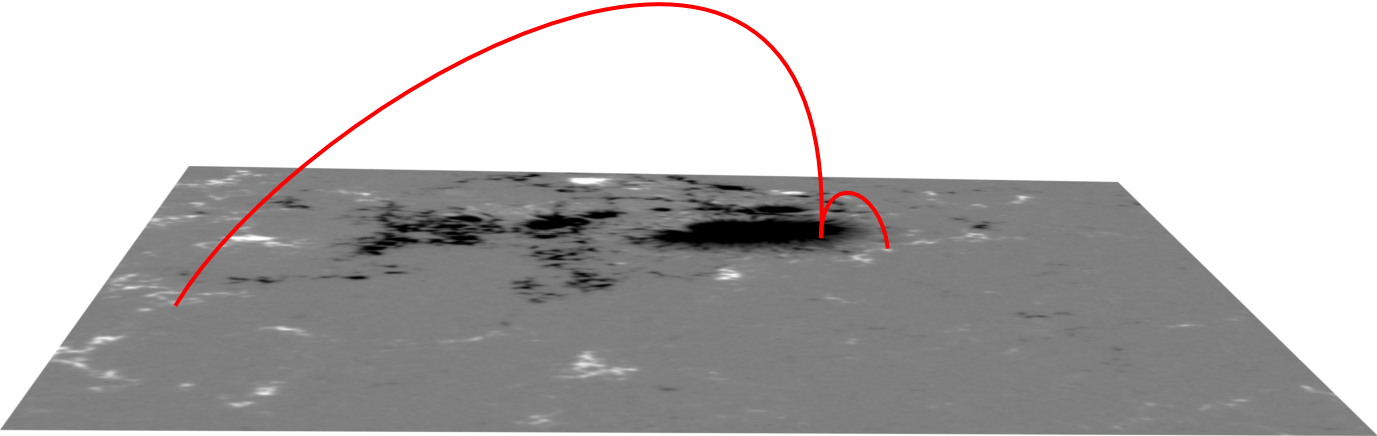}
\caption{Three-dimensional renderings of the small and large scale lines of force (in red) and perspective views of the HMI magnetogram, as viewed from the east (top) and from the south (bottom).  The field of view is 180 by 180\arcsec.
}
\label{3D}
\end{figure}

\subsection{Mass motions beyond the penumbral brightening}
We note that the ``tails'' of the structures in Fig. \ref{cutsA} are slightly inclined with respect to the direction of position, $x$, which shows that we have motions or, alternatively, a propagation of the brightening towards the outer footpoint of the loop; this shows better in the 304 and the higher temperature bands, where the cadence is higher.  For the main event the apparent motion had a velocity on the plane of the sky of the order of 100\,km\,s$^{-1}$ in all bands except for 94\,\AA, where it reached $\sim200$\kms. We note, for reference, that the speed of sound for a million-degree plasma is about 150\kms.

\begin{figure}[h!]
\centering
\includegraphics[width=.95\hsize]{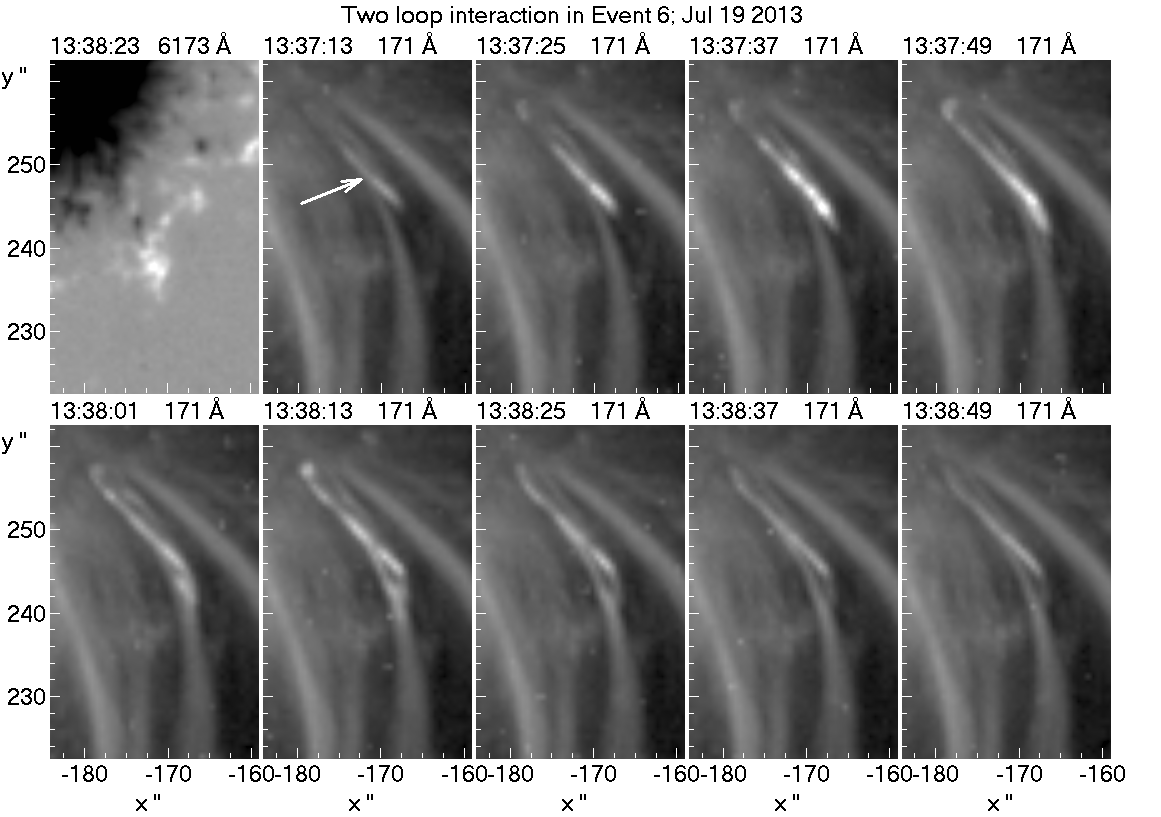}
\caption{Two-loop interaction and mass flows in Event 6 in 171\ang\ AIA images; an HMI magnetogram is given for reference. The arrow points to the intersection of the projections on the plane of the sky of the high and low loops.
 The full sequence of images in the 1600, 1400, 304, 171 and 335\,\AA\ bands is given in the movie attached to Fig. \ref{movie02}.}
\label{Event6}
\end{figure}

A closer look reveals mass motions beyond the penumbral brightening. This can be seen in the 304\ang\ image of Fig. \ref{overview}, near the SW end of the brightening and in the movie attached to Fig. \ref{movie01}; it is clearer in Fig. \ref{flows} where we show images of the intensity RMS in five spectral bands, together with a magnetogram. The flows are more prominent in the 304\ang\ band, less prominent in the 171\ang\ band and even less at 335\ang. They are  hardly visible in the 1400\ang\ IRIS JSs and completely invisible at 1600\ang. 

We considered the possibility that the flows, at least in part, occurred not in the flaring penumbral loop but in a different one. Indeed, the AIA images show several large loops starting in the penumbra and ending far away; as a matter of fact, there is a large loop very near our flaring loop. This was confirmed by linear force-free extrapolation \citep{1981A&A...100..197A} of the magnetic field based on the HMI magnetogram at 13:18:53 UT. The value of the force-free parameter, $\alpha$, was determined by minimizing the distance between the computed magnetic field lines, projected on the plane of the sky \citep{2004ApJ...616L.175N} and the loops starting near the inner footpoint of the brightening in the AIA 171\ang\ image at 13:18:47 UT; this gave $\alpha=-0.4$. The results show that indeed, there was a large-scale loop that projected very near the small flaring loop (Fig. \ref{FieldLines}).

The projections on the plane of the sky of the lines of force of the small and the large loop, plotted on the magnetogram of Fig. \ref{flows}, are very close to each other near the brightening, considering the limitations of the linear force-free extrapolation. Three-dimensional views of these lines, together with the magnetogram in perspective, are shown in Fig. \ref{3D}. We note that, according to the extrapolation, the angle between the loops and the line of sight at the position of the spectrograph slit are 30 and 42\degr\ for the large and small loop respectively, while their distance is 800\,km. From its middle to the outer footpoint the small loop is nearly perpendicular to the line of sight ({\it cf\/}. Fig. \ref{Loops}), hence the velocities quoted above should be close to the true velocities. 

Our hypothesis for loop interaction is corroborated by the sequence of images of the homologous event 6, shown in Fig. \ref{Event6}. In the first frames we can see a large-scale loop, whose projection on the plane of the sky intersects that of the lower flaring loop. In the course of the event, an intense brightening forms near the intersection (frames at 13:37:37 and 13:37:49 UT), while in the next frame enhanced emission appears in the large-scale loop and moves to the south. The full sequence of frames in the AIA wavelength bands together with HMI magnetograms is given in the movie attached to Fig. \ref{movie02}.

\begin{figure*}
\centering
\includegraphics[width=.8\hsize]{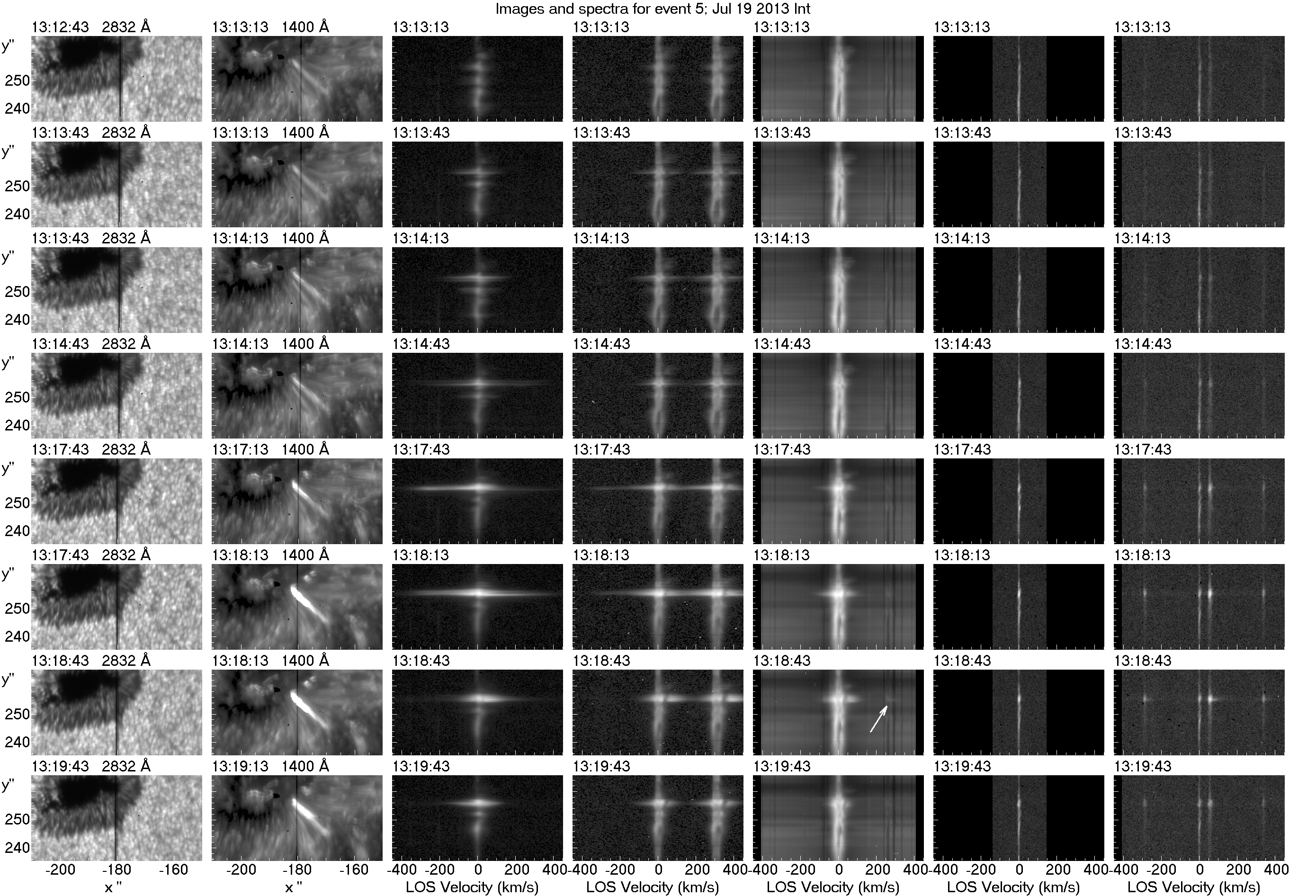}
\caption{IRIS SJ images in the 2832 and 1400\,\AA\ bands and spectra of {Si \sc iv} 1393.8\,\AA, \Cb\ doublet (centered at the 1334.6\,\AA\ line), \Mg\ k, 1351.7\,\AA\ {Cl \sc i} and the {O \sc i} 1355.6\,\AA\ -- {C \sc i} 1355.8\,\AA\ lines during the main event. Missing SJ images, due to their lower cadence, are replaced by the previous ones. The arrow points to a weak emission from the \Mg\ triplet line. The wavelength is expressed in terms of the line-of-sight velocity, with the origin at the line center. The images and the spectra in each column are displayed with the same intensity scale. For a more complete sequence see the movie attached to Fig. \ref{movie03}.
}
\label{spectra}
\end{figure*}
\begin{figure*}
\centering
\includegraphics[width=.9\hsize]{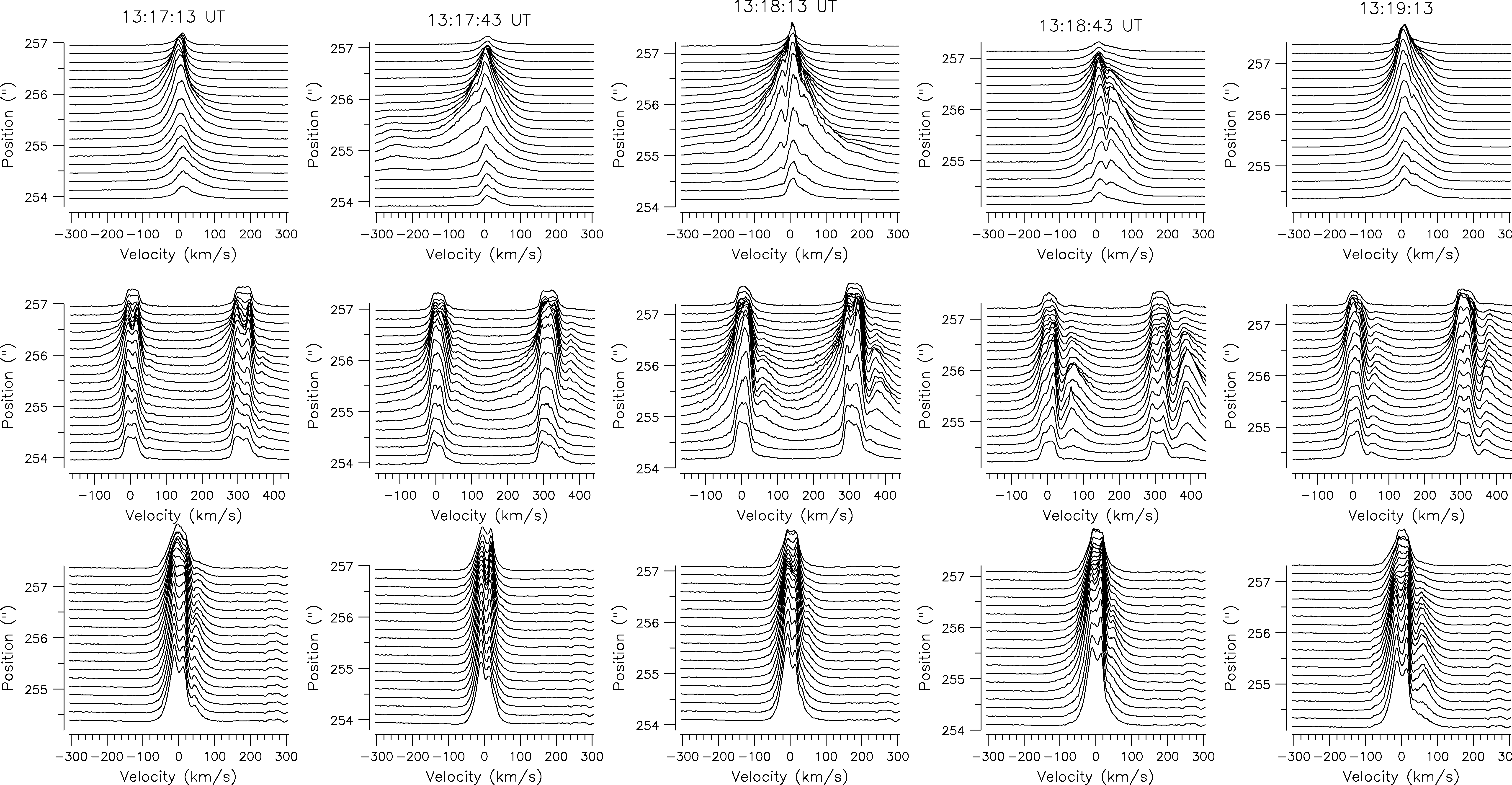}
\caption{Profiles of {Si\,\sc iv} 1393.8\,\AA\ (top), {C\,\sc ii}  1334.6\ang\ (center) and {Mg\,\sc ii} k (bottom) lines during the main event. Time is marked above each {Si\,\sc iv} spectrum. The intensity scale is the same for all profiles of the same line.
}
\label{SpecProf}
\end{figure*}

\section{Spectral analysis}

\subsection{Overview of the main event}\label{SpecOver}
During the main event the IRIS slit was near the inner footpoint of the loop, very close to the intensity peak ({\it cf.} Fig \ref{overview}). In the course of the event the slit drifted slowly towards the footpoint, by $\sim0.22$\arcsec\,min$^{-1}$; the feature being inclined by 44\degr\ with respect to the slit, this results in an upward apparent motion of its signature in the spectra at the same rate.

Figure \ref{spectra} shows a sequence of images and spectra for the main event which covers the spike as well as a minor maximum at 13:14:43 UT ({\it cf.} Fig. \ref{timprof}). In addition to the  {Si\,\sc iv}, {C\,\sc ii,} and {Mg\,\sc ii} k lines, we give spectra of the 1351.7\,\AA\ {Cl\,\sc i} line as well as of the {O\,\sc i} 1355.6\,\AA\ and the {C\,\sc i} 1355.8\,\AA\ lines. We have used a velocity rather than a wavelength scale, with the zero determined from the average Doppler shift of quiet regions along the slit. A more complete sequence is given in the movie attached to Fig. \ref{movie03}, where we give all spectra from 13:12 to 13:22, as well as images in the 304\,\AA\ AIA band and IRIS SJs. Unlike the 1400\ang\ SJs, the spectra do not suffer from saturation. A selection of line profiles is plotted in Fig. \ref{SpecProf}, which will be discussed in the next subsection.

Apart from the time variation of intensity, the most striking feature in the spectra is the very large width of the {Si \sc iv} and \Cb\ lines near the peak of the event. The thermal width being quite small, of the order of 6-7\kms,\ and the instrumental width half of that, the broadening is certainly of non-thermal origin. For the {Si \sc iv} lines in particular, which are optically thin as evidenced from the comparison of the intensities of the doublet, the profiles reflect with high accuracy the velocity distribution of non-thermal motions. Of the two {Si \sc iv} lines, the one at 1393.8\,\AA\ is in a region clear of blends and for this reason we will start our description from this particular line. The \Cb\ and \Mg\ lines are optically thick, as evidenced from their double-peaked profiles. 

With profiles of such shape it is not very meaningful to compute average line-of-sight (LOS) velocities and widths. However, it is interesting to note that the spectral maximum remains near 0\kms\ throughout the event, while the wings change dramatically. Compared to the pre-event situation at 13:13:13 UT, we have strong blue wing emission in the spectrum just before the peak, at 13:17:43 UT. There is a noticeable gradient across the brightening, with the lower part exhibiting a stronger red wing than the upper. An intense red wing is added to the spectrum 30\,s later, while the intensity of the blue wing gradually diminishes. We note that the orientation of the loop with respect to the line of sight, deduced from our extrapolation of the magnetic field, is such that redshifts correspond to downflows. 

The wing emission is so much extended that it is well above the background even beyond $\pm400$\kms. Moreover, there is a discrete blue shifted component at  $-250$\kms, with a FWHM of $\sim80$\kms, well visible both in the {Si \sc iv} and the \Cb\ lines; this is stronger in the 13:17:43 spectrum, but still detectable in the next. With the magnetic lines of force forming an angle of 42\degr\ with the line of sight (see previous section), this blue shift translates to 340\kms\ along the loop and may be part of the mass motions we detected in AIA images.
The extended wings  disappeared in the spectrum at 13:18:43 UT, where the red wing was stronger than the blue, while at 13:19:43 the blue wing was stronger again. The spectra during the weak peak at 13:14:13 (Fig. \ref{timprof}) also show wide emission in both wings, with the blue wing predominating before and after the maximum.

\begin{figure}
\centering
\includegraphics[width=\hsize]{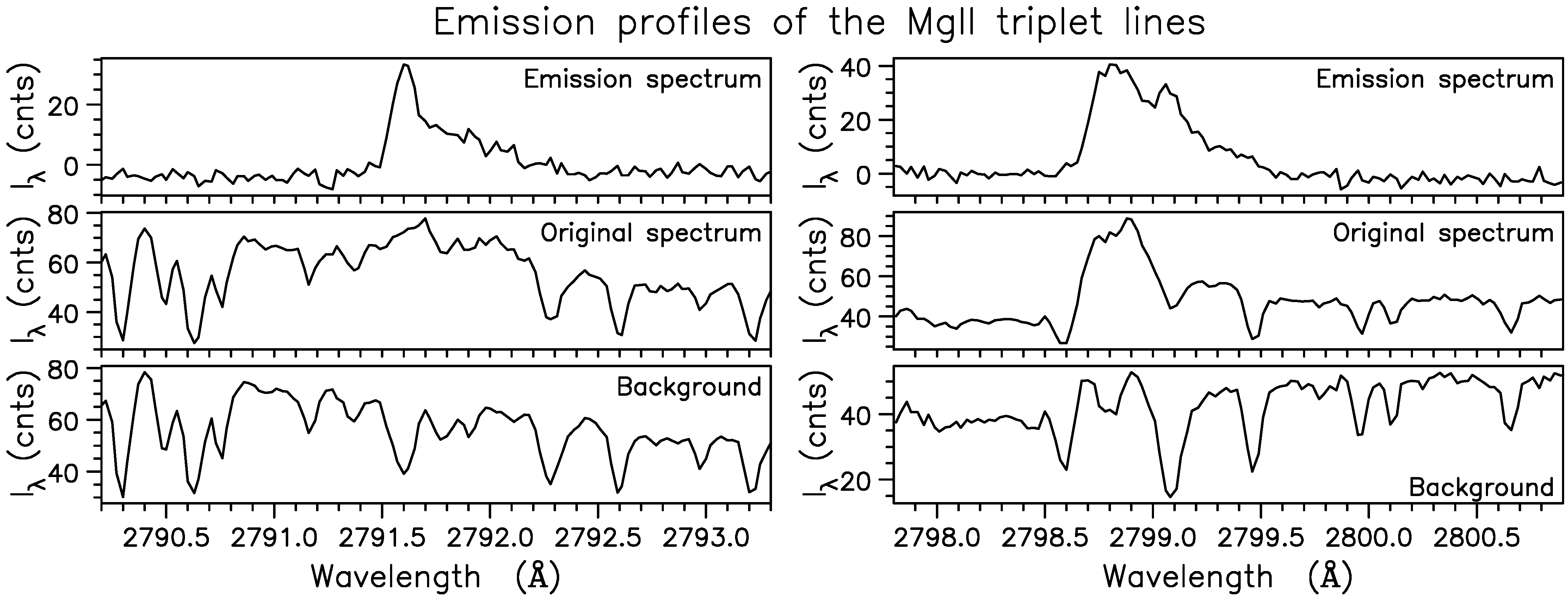}
\caption{Emission profiles of the {Mg\,\sc ii} triplet lines at  2791.6\,\AA\ (left) and 2798.8\,\AA\ (right) at 13:18:43 UT. The plots show the background spectrum (lower row) and the spectrum through the feature before (middle row) and after (top row) background subtraction.  
}
\label{EmProf}
\end{figure}

Although the k and h lines show wing emission, this is not as extended as in the {Si\,\sc iv} and the {C\,\sc ii} lines. In its red wing a weak emission appears at 13:18:43 UT (arrow in Fig. \ref{spectra}), barely visible in the previous and next frames. This comes from the 2798.8\,\AA\ {Mg\,\sc ii} triplet line \citep[a blend of the 2798.75 and 2798.82\,\AA\ lines, see][]{2015ApJ...806...14P}; it is better visible~in the top frame of the right panel of Fig. \ref{EmProf}, where the background spectrum has been subtracted. There is also emission from the nearby Mn\,{\sc i} line at 2799.1\,\AA, as well as from the other triplet line at 2791.7\,\AA\ (left panel of Fig. \ref{EmProf}). In all cases the profiles show extended red wings, as do the simultaneous {Si\,\sc iv}, {C\,\sc ii,} and  {Mg\,\sc ii} profiles. 

The spectra in the last two columns of Fig. \ref{spectra} show that we have event-associated  emission in the weak, low temperature lines of {Cl\,\sc i} at 1351.7\,\AA, {O\,\sc i} at 1355.6\,\AA,\ and {C\,\sc i} at 1355.8\,\AA. We have weaker emission in the 1354.3\,\AA\ and 1357.1\ang\  {C\,\sc i} lines, located at $-290$ and 340\kms\ with respect to the {O\,\sc i} line. 

The strongest of these is the {Cl\,\sc i} line, which shows slightly tilted spectra, evidence of a gradient in the LOS velocity  across the feature, which did not exist before the event. Such tilts are usually interpreted in terms of rotation; if this is the case, the measured LOS velocity gradient of $\sim1$\kms/\arcsec\ translates to a clockwise rotation with an angular velocity of $\sim3$\degr\,min$^{-1}$, for a feature inclined by 45\degr\ with respect to the slit. This, in turn, implies a twisting or, most probably, an untwisting of the magnetic flux tube by about 15\degr\ during the lifetime of the event. Interestingly, the sense of the inferred rotation (clockwise) is consistent with
the helicity sign of the magnetic field configuration as resulting from the magnetic field extrapolation of Section 3.3.

Similar LOS velocity gradients are present in the other weak lines of this spectral range; however, they are not limited to our particular feature, but they appear elsewhere in the spectra (e.g. near the position 248\arcsec). An additional interesting feature is that the {Cl\,\sc i} line and the {C\,\sc i} 1355.8\,\AA\ line show extended red wings in the spectra at 13:48:43 UT, like the {Mg\,\sc ii} triplet lines. 

\subsection{Absorption features in spectral profiles}
\begin{figure}
\centering
\includegraphics[width=\hsize]{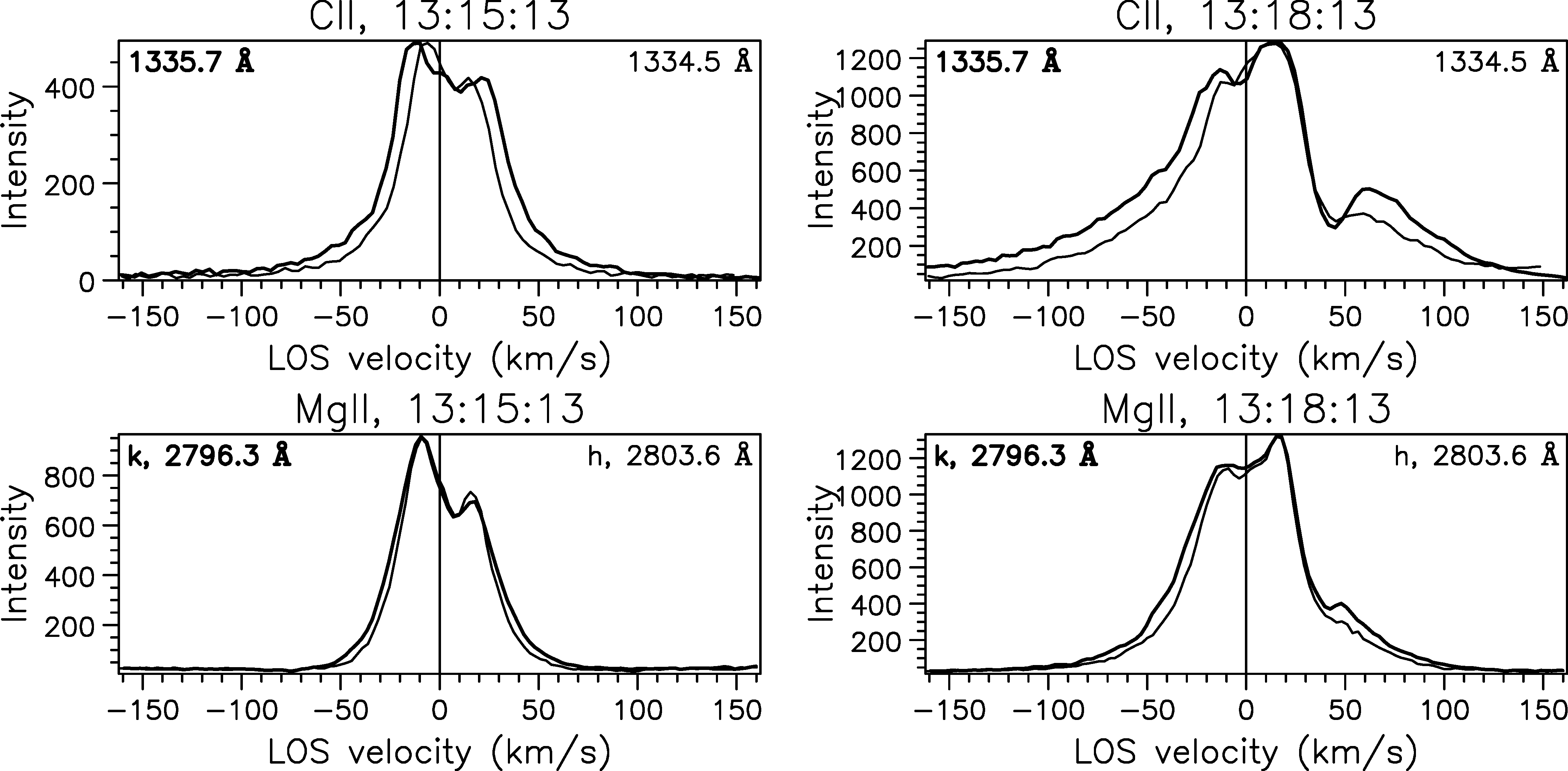}
\caption{Profiles of {C\,\sc ii} (top) and {Mg\,\sc ii} (bottom) lines without (left) and with (right) absorption features.  The intensity of each profile is scaled so that the plot covers its full range from the bottom to the top. Thick lines show the profiles of the strongest line of the doublet, to which the intensity scale also refers.
}
\label{SpecProFDoublet}
\end{figure}

In addition to very broad wings, some spectra show absorption features, more prominent in the red wing of the {C\,\sc ii} lines (see spectra near 13:18:13 UT in Fig. \ref{spectra}); they are better seen in the selected profiles of Fig. \ref{SpecProf}. That these features are not due to blends is proven by the fact that they appear at the same position in both lines of the doublet pairs. We can also exclude the possibility that they are due to a peculiar form of the velocity distribution function, rather than due to absorption, by comparing the profiles of the {C\,\sc ii} and {Mg\,\sc ii} doublet lines (Fig. \ref{SpecProFDoublet}). As expected for optically thick lines, the stronger line of the doublet is wider and this is symmetric with respect to the line core (profiles at the left of the figure). However, in the profiles at the right, the strong line is broader than the weak one in the blue wing  but not in the red, proving that something is absorbing there. We add that the absorption affects the strong line of the {C\,\sc ii} and {Mg\,\sc ii} doublets more than the weak one, which indicates that the absorption is not optically thin. We cannot provide similar arguments for {Si\,\sc iv}, because the lines are optically thin, but we may presume by analogy that we have a similar situation.

As we follow the evolution of the absorption feature in the {C\,\sc ii} 1334.6\ang\ spectra (see Fig. \ref{spectra} and the movie attached to Fig. \ref{movie03}), we note that it first appeared with a redshift of $\sim$55\kms\ at 13:13:43 UT, in association with the weak peak at 13:14:13 (Fig. \ref{timprof}); it became weaker 30\,s later and disappeared at 13:14:43. The absorption reappeared\index{red}\index{red} at 13:17:13 UT (Fig. \ref{SpecProf}), during the rise phase of the impulsive peak and was visible up to 13:20:43 UT in its late decay phase, being stronger at 13:18:43 UT, 30\,s after the impulsive peak. During this time interval, the velocity shift decreased slowly to $\sim40$\kms. The spectral width of the absorption was much more narrow than the emission profile, $\sim18$\kms; this is considerably larger than the thermal width (Section \ref{SpecOver}) and comparable to the width of the quiet sun emission profiles.

The absorption was less prominent in the {Mg\,\sc ii} lines (Figs. \ref{spectra}, \ref{SpecProf} and the movie attached to Fig. \ref{movie03}), where it was detectable for a shorter interval, from 13:17:43 to 13:19:43 UT; its shift from the line center was smaller than for the {C\,\sc ii} lines, being 38\kms\ at 13:18:43 UT, compared to 42\kms\ in {C\,\sc ii}. Its width was also smaller than in {C\,\sc ii}, $\sim15$\kms. In the {Si\,\sc iv} profiles which, as mentioned previously, are optically thin and of identical shape for the two components of the doublet, the absorption appeared on two spectra only: the one at the impulsive peak and the one after that (13:18:43 UT), being stronger in the latter; its shift was even smaller than in {Mg\,\sc ii}, 30\kms\ and so was its width, $\sim10$\kms. During  the impulsive peak, the  {Si\,\sc iv} spectrum showed a second absorption feature, at $-12$\kms\ (best seen in Fig. \ref{SpecProf}), which is not visible in any other line. This is stronger in the lower part of the bright structure, with some traces left in the next spectrum.

In addition to the varying Doppler shift, the absorption features show tilted profiles near the impulsive peak. In both the {C\,\sc ii} and the {Mg\,\sc ii} lines the tilt jumped from nearly zero to $\sim6$\kms/\arcsec\ 30\,s before the impulsive peak; in the {Si\,\sc iv} lines we could measure the tilt only in the spectrum after the peak and this gave a smaller value of $\sim2$\kms/\arcsec. These tilts are considerably larger than those measured from the {Cl\,\sc i} line which, moreover, did not exhibit the Doppler shift of the absorption features seen in the strong lines; this is not surprising since we expect the {Cl\,\sc i} line to form in a different region. The tilts, however, are in the same sense. Interpreted in terms of rotation, they translate to 18\degr\,min$^{-1}$ for {C\,\sc ii} and {Mg\,\sc ii} and 6\degr\,min$^{-1}$ for {Si\,\sc iv}, compared to 3\degr\,min$^{-1}$ for {Cl\,\sc i}. 

Absorption features in IRIS spectra have been reported by \cite{2014Sci...346C.315P} from {Ni \sc ii} in the blue wing of the {Si\,\sc iv} 1393.8\ang\ line and by \cite{2015ApJ...811...48Y} from the {Si\,\sc iv} lines themselves. \cite{2014Sci...346C.315P} interpreted them in terms of co-existing cool and hot material, while \cite{2015ApJ...811...48Y} in terms of self-absorption, which implies absorption from the outer layers of the emitting structure. 

In any case, the absorption in {Si\,\sc iv} and, in our case, in {C\,\sc ii} and {Mg\,\sc ii} as well, indicates the presence of material with lower source function between the emitting structure and the observer; in addition, the spectrum of the absorbing plasma is redshifted and has a smaller Doppler width compared to that of the emitting plasma. As the most likely source of this material we consider the plasma in the upper loop discussed in Sect. 3.3. In this scenario the redshift of the absorption is, at least in part, due to a descending motion of the upper loop which thus approaches the lower loop, already energized. It is possible that the interaction of the two loops enhances the on-going slow energy release and gives rise to the spiky component of the emission. The observed tilts of the spectral lines and their time evolution are naturally interpreted in terms of the untwisting of flux tubes, as a result of the interaction. Finally, the strong interaction of the two loops induces mass motions in the upper loop, as discussed in Sect. 3.3.

\subsection{Density diagnostics}
We estimated the electron density from the intensity ratio of the {O\,\sc iv} lines at 1399.78 and 1401.16\ang, based on the computations of \cite{2016A&A...594A..64P}. Although these lines are weak on the quiet sun, they give a measurable signal at the brightening. The main observational problem is that, due to the very wide wings of the nearby \Si\ line at 1402.80\ang, we could not measure the intensities of the {O\,\sc iv} lines near the peak of the event. Before and after the maximum we obtained an intensity ratio of 0.4, which corresponds to an electron density of $N_e=7\times10^{11}$\,cm$^{-3}$. This value is one order of magnitude higher than the one obtained by \cite{2016A&A...594A..64P} for an active region loop and one order of magnitude lower than what they derived for the impulsive phase of a flare from the ratio of \Si\ intercombination lines. Interestingly, \citet{2016ApJ...832...77D} found plasma-bulk electron densities a few times $10^{11}$\,cm$^{-3}$ from the analysis of transition region explosive events in the {O\,\sc iv} doublet we also used.

We consider our value of electron density as indicative only, since we may be quite far from ionization equilibrium in our fast-evolving brightening.  Such departures could lead to even higher densities \citep{2013ApJ...767...43O}.

\subsection{Line profiles from other events}
Unfortunately, the spectrograph slit did not cross event 6. During the other brightenings the slit was near the outer footpoint for event 1 and near the middle of the loops for events 2, 3 and 4 (see Fig. \ref{others}). These were weak events and none of them had profiles as broad as event 5. In event 1 the \Si\ red wing was stronger and detectable up to 200\kms, while in the others the profiles were more symmetric and the wings did not extend beyond $\pm100$\kms. This difference could be due to a geometric effect, with the slit being near the middle of the loops and the line of sight being nearly perpendicular to the loop axes; in such a case, it indicates that non-thermal velocities are much larger along the loop than in the perpendicular direction. No absorption features were detected in these events in \Si\ or any other line. 

\section{Summary and conclusions}
The small-scale penumbral brightenings, studied in this work using simultaneous IRIS and SDO observations, revealed a number of interesting aspects. In all cases, they involved low-lying loops of sub-arcsecond width, with one footpoint near the umbra-penumbra boundary and the other in nearby opposite polarity regions, at a distance of 17-20\arcsec. Three of them were practically at the same location and the other three were very near the former. We found nothing peculiar in the photospheric magnetic field at the location of the brightenings

We analyzed in detail the brightest event. The intensity structure was very similar in all wavelength bands, with a very bright maximum near the inner footpoint and other small-scale structures along the loop, more prominent in the low-temperature channels. It had an impulsive peak, lasting for about 2 min and consisting of two to three shorter components. Time profiles were very similar in all channels, with a few seconds delay in the low-temperature channels with respect to the high-temperature ones, which may indicate that the energy release occurred in the upper transition region or the low corona. The impulsive peak was superposed on a 10-min long slowly varying background, while the event occurred in a region that had enhanced brightness during the entire observing sequence, which shows that there was continuous low-level activity there. In contrast to other small scale events we have studied previously \citep{2013A&A...556A..79A, 2015A&A...582A..52A}, no post-burst phase was detected.

During the event the IRIS spectrograph slit was near the bright inner footpoint. Close to the maximum, the \Si\ and \Cb\ profiles were very broad, with the former standing clearly above the continuum even beyond $\pm400$\kms. These are among the broadest profiles reported from IRIS observations \citep{2014Sci...346C.315P, 2014ApJ...797...88H, 2015ApJ...813...86I, 2016ApJ...829...35W}. The \Si\ lines being optically thin, with a thermal width much smaller than the observed width, their profiles reproduce the distribution of the non-thermal velocities. Thirty seconds before the maximum of the event we had strong blue-wing emission (upflows), indicative of chromospheric evaporation; there was also a discrete component at $-250$\kms, visible both in the \Si\ and the \Cb\ lines and probably associated to the mass motions we detected in the AIA images. Subsequently the red wing dominated, while near the end of the impulsive phase the blue wing became stronger again. 

The \Mg\ lines also showed enhanced wing emission, but less extended than the {Si\,\sc iv} and the {C\,\sc ii} lines. In addition, we detected emission from narrow weak lines formed at low temperatures: the {Mg\,\sc ii} triplet lines, the Mn\,{\sc i} 2799.1\,\AA\ line, the 1351.7\,\AA\ {Cl\,\sc i} line, the {O\,\sc i} line at 1355.6\,\AA,\ and the {C\,\sc i} lines at 1355.8, 1354.3 and 1357.1\ang, which all showed enhanced red wing emission. The strongest line in this group, that of {Cl\,\sc i}, had slightly tilted spectra, evidence of a gradient in the LOS velocity across the feature, which appeared during the event. Interpreted in terms of rotation, the tilts imply an untwisting of the magnetic flux rope by about 15\degr\ during the duration of the event;  the sense of the inferred rotation (clockwise) is consistent with the helicity sign of the magnetic field configuration deduced from the magnetic field extrapolation.

Narrow absorption features appeared in the \Cb, \Mg,\ and \Si\ profiles around the main maximum of the event as well as near a secondary maximum 5 min before the main. In \Cb\ they were 18\kms\ wide and redshifted with respect to the main emission profile by 55\kms\ at 13:13:43 UT to 40\kms\ at 13:18:43 UT. They were less prominent in {Mg\,\sc ii} with a smaller Doppler shift, 38\kms\ and smaller width, 15\kms. In \Si\ the shift was even smaller, 30\kms\ and so was the width, 10\kms; moreover,  a second absorption feature appeared in these lines during  the impulsive peak, blue-shifted at $-12$\kms. The absorption profiles showed tilts, of the same sense but larger than those measured in \Cl.

Absorption features in \Cb\ and \Mg\ are reported here for the first time. \cite{2014Sci...346C.315P} reported absorption from {Ni \sc ii} in the blue wing of the {Si\,\sc iv} 1393.8\ang\ line and attributed it to co-existing cool and hot material, while \cite{2015ApJ...811...48Y} reported absorption from the {Si\,\sc iv} lines themselves and interpreted it as self-absorption.

As the most likely source of the absorption we consider the plasma in a quiescent loop located between the flaring loop and the line of sight. Indeed, the AIA images show several loops starting in the penumbra; moreover, the projection on the plane of the sky of one large loop practically coincided with that of the brightening near its inner footpoint, while the other footpoint of the large loop was far away. This was verified by linear force-free extrapolation of the photospheric magnetic field, which reproduced in a satisfactory way, given the limitations of such extrapolations, the shapes of both the low flaring loop and the high quiescent loop for the main event. The two-loop interaction was more clear in the homologous event 6 (Fig. \ref{Event6} and the movie attached to Fig. \ref{movie02}), for which we had no spectral information. The blue-shifted absorption observed in \Si\ could originate in a third loop.

\begin{figure}[ht]
\centering
\includegraphics[width=.8\hsize]{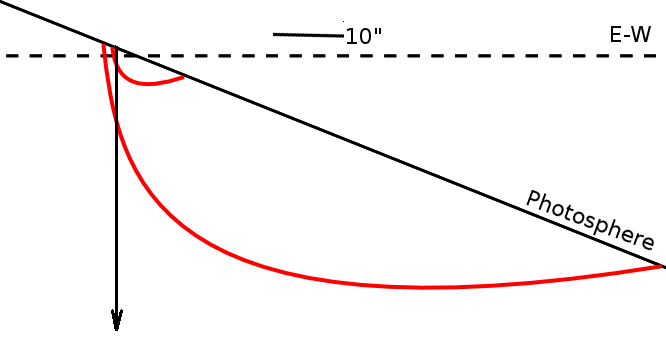}
\caption{Projections of the high and low loops from our linear force-free computations, on a plane defined by the line of sight and the solar EW direction. The arrow points towards the observer and is drawn at the position of the spectrograph slit.
}
\label{Loops}
\end{figure}

The geometry (Fig. \ref{Loops}) is such that part of the observed redshift of the absorption can be attributed to downflows  and another part to descending motion of the entire upper loop, which thus approaches the lower loop, already energized. It is therefore possible that the interaction of the two loops enhances the on-going slow energy release, manifested by the gradual increase of the intensity and gives rise to the spiky component of the emission as well as to mass motions in the upper loop. 

Configurations with more than one loop appear to be common in various flares: from small spiky events 
\citep{2004A&A...420..351K}, to large, long-duration events \citep{2017SoPh..292....3G}. Reconnection is expected to change the topology of the region quite dramatically, which is not the case with our events. This indicates that here we have a slow reconnection process that preserves the geometry and is responsible for the continuous low-level activity, while the interaction of the two loops enhances the energy release process and produces the brightening and mass flow without changing the geometry of the configuration. Moreover, we are dealing here with events close to the spatial resolution limit of the corresponding instruments. This is possibly an additional factor preventing us from seeing clear manifestations of magnetic reconfigurations during our event.

Our study highlights that energy release events which may have a bearing on coronal heating do not necessarily involve small spatial scales only. They may well implicate interacting loops of different sizes as in the case reported here, where a small loop interacted with a large-scale loop whose footpoint separation was similar to the typical spatial scale of the active region. Moreover, our study shows that two-loop interactions are relevant not only to proper flares, but also to smaller scale energy release, suggesting scale-invariant processes. It would be interesting to investigate in the future how frequent such small-scale two-loop interactions are.

\begin{acknowledgements}
The authors gratefully acknowledge use of data from the IRIS and SDO (AIA and HMI) databases as well as images from the Helioviewer site. IRIS is a NASA small explorer mission developed and operated by LMSAL with mission operations executed at NASA Ames Research center and major contributions to downlink communications funded by ESA and the Norwegian Space Centre.

\end{acknowledgements}

\begin{appendix}
\section{Supplementary material}
In this appendix we present the movies discussed in the text.
\begin{figure*}
\centering
\includegraphics[width=.8\hsize]{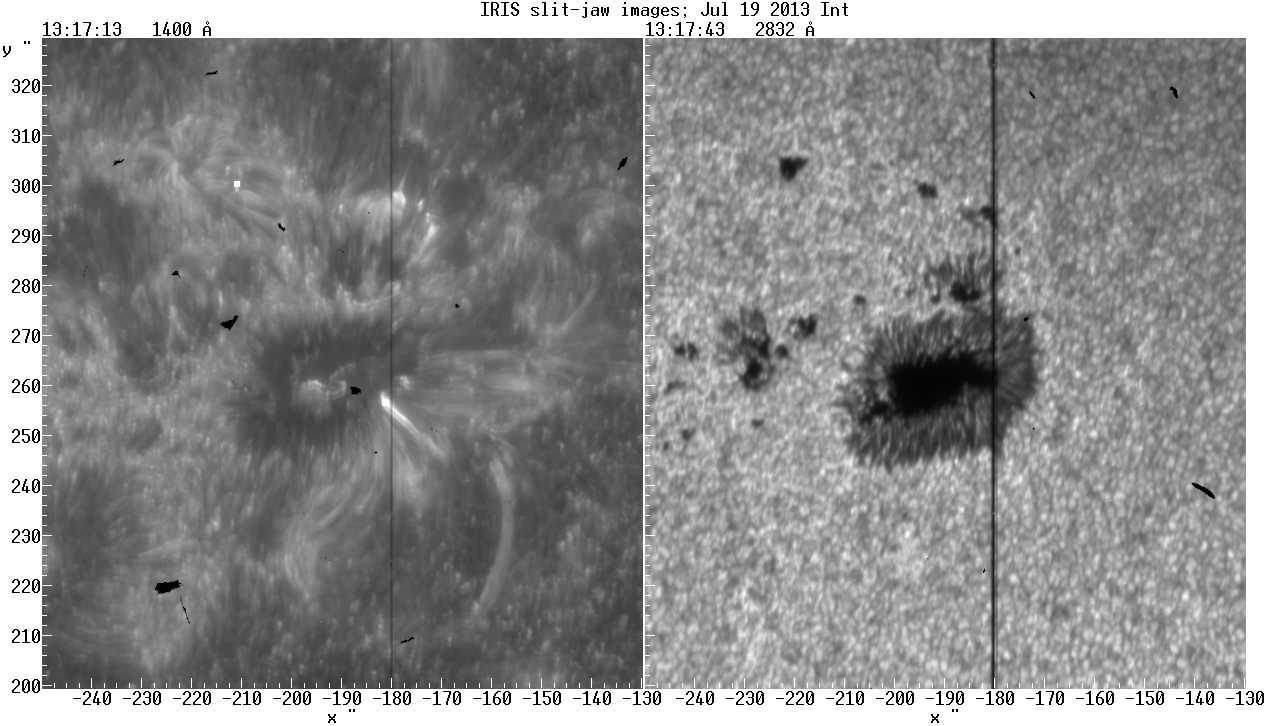}
\caption{A frame of movie 1, which shows the full set of IRIS SJ images in the 1400\ang\ and 2832\ang\ bands.
}
\label{movie01}
\end{figure*}

\begin{figure*}
\centering
\includegraphics[width=.8\hsize]{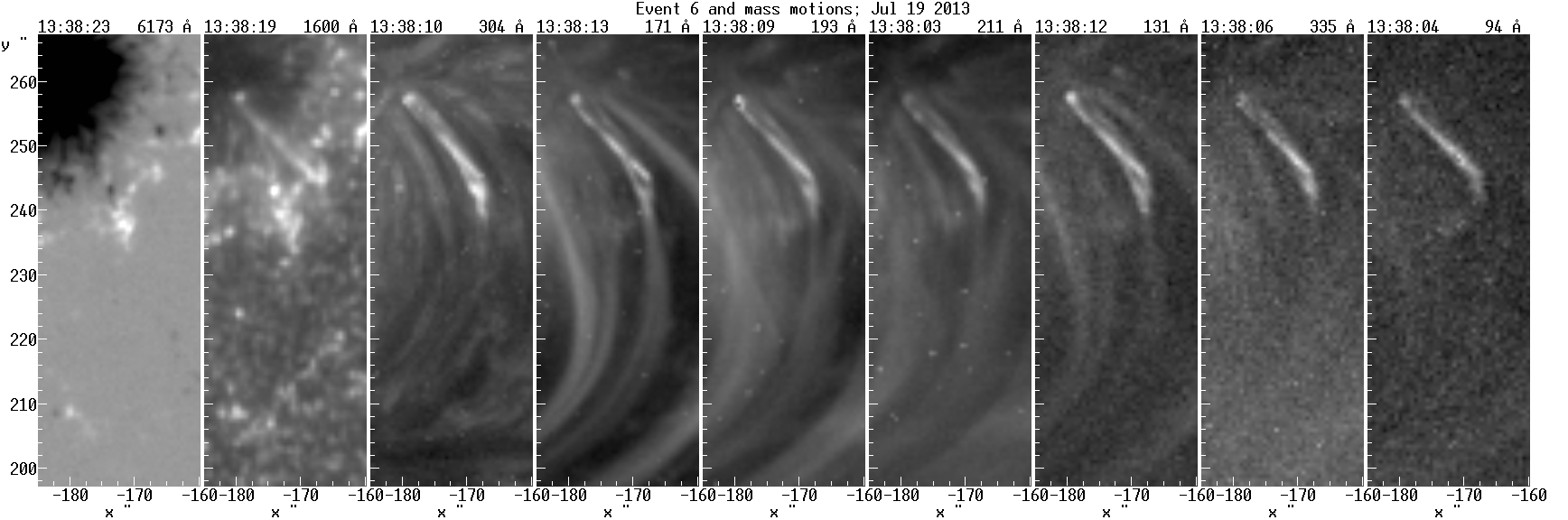}
\caption{A frame of movie 2, which shows the interaction of the two loops in event 6. From left to right: HMI magnetogram and AIA images in the 1600\ang, 304\ang, 171\ang 193\ang, 211\ang, 131\ang, 135\ang,\ and 94\ang\ bands.
}
\label{movie02}
\end{figure*}

\begin{figure*}
\centering
\includegraphics[width=.8\hsize]{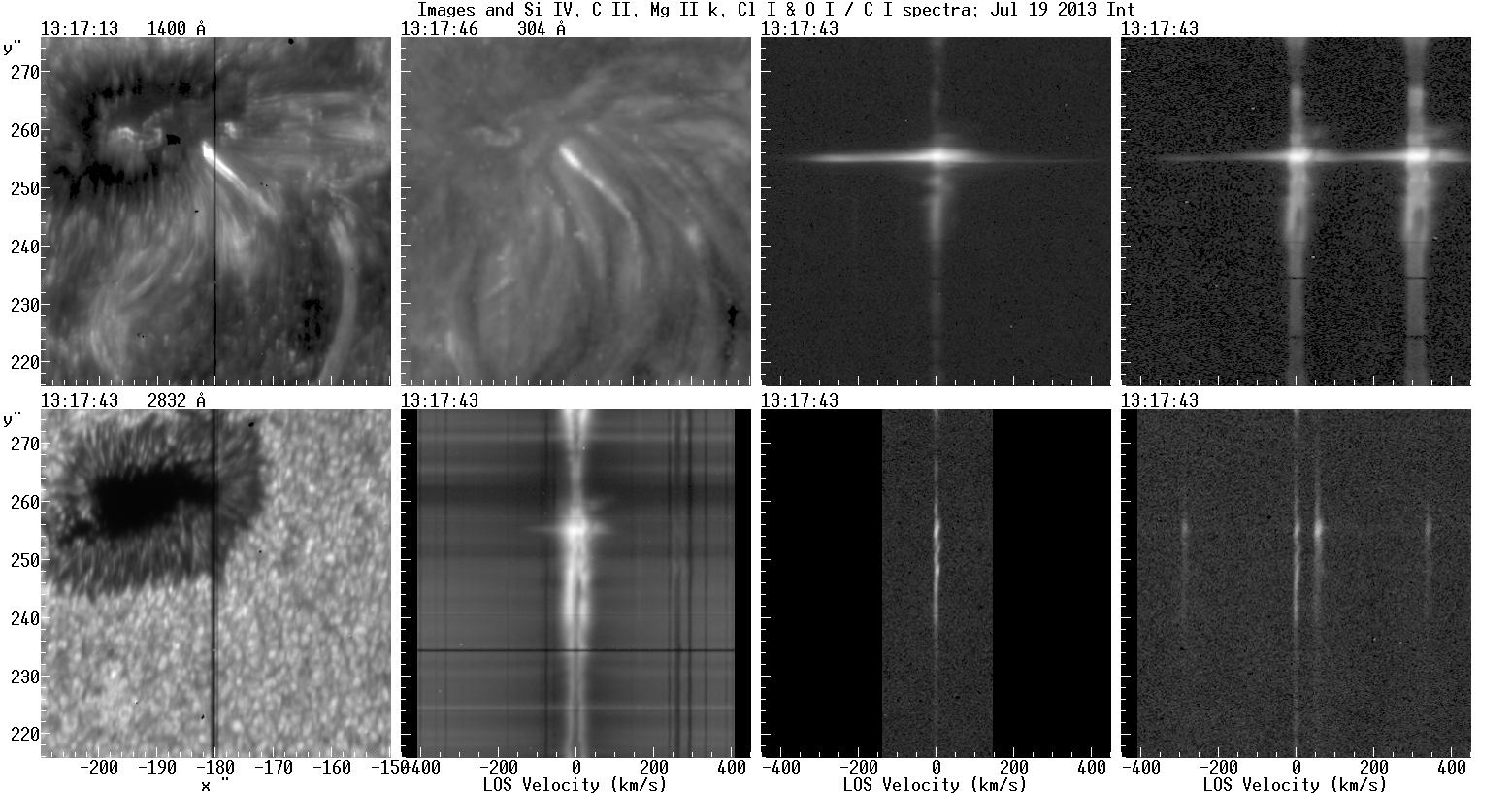}
\caption{A frame of movie 3, showing a sequence of IRIS SJ images in the 1400\ang\ and 2832\ang\ bands, AIA images in the 304\ang\ band and IRIS spectra in the {Si\,\sc iv} 1393.8\,\AA, \Cb\ doublet (centered at the 1334.6\,\AA\ line), \Mg\ k, 1351.7\,\AA\ {Cl \sc i} and the {O \sc i} 1355.6\,\AA\ -- {C \sc i} 1355.8\,\AA\ lines, in the course of the main event The wavelength scale of the spectra is in units of line-of-sight velocity.
}
\label{movie03}
\end{figure*}

\end{appendix}

\printindex

\end{document}